\documentclass[journal]{IEEEtran}
\IEEEoverridecommandlockouts
\usepackage{algorithmic}
\usepackage{array}
\usepackage[caption=false]{subfig}
\usepackage{textcomp}
\usepackage{stfloats}
\usepackage{url}
\usepackage{verbatim}
\usepackage{graphicx}
\usepackage{cite}
\usepackage{amsmath,amssymb,amsfonts}
\usepackage{xcolor}
\usepackage{overpic} 
\usepackage{cuted}
\usepackage[linesnumbered,ruled,noend]{algorithm2e}
\usepackage{bm}
\usepackage{booktabs}
\def\BibTeX{{\rm B\kern-.05em{\sc i\kern-.025em b}\kern-.08em
    T\kern-.1667em\lower.7ex\hbox{E}\kern-.125emX}}

\begin{document}
\title{When UAV Swarm Meets IRS: Collaborative Secure Communications in Low-altitude Wireless Networks}

\author{
Jiahui Li,
Xinyue Liang,
Geng~Sun,~\IEEEmembership{Senior Member,~IEEE,}
Hui~Kang,
Jiacheng Wang, \\
Dusit Niyato,~\IEEEmembership{Fellow,~IEEE,} 
Shiwen Mao,~\IEEEmembership{Fellow,~IEEE,}
Abbas Jamalipour,~\IEEEmembership{Fellow,~IEEE}

  \thanks{
  
  \par Jiahui Li, Xinyue Liang, and Hui Kang are with the College of Computer Science and Technology, Jilin University, Changchun 130012, China (E-mails: lijiahui@jlu.edu.cn; liangxy2121@mails.jlu.edu.cn; kanghui@jlu.edu.cn). 
  
  \par Geng Sun is with the College of Computer Science and Technology, Jilin University, Changchun 130012, China, and also with the Key Laboratory of Symbolic Computation and Knowledge Engineering of Ministry of Education, Jilin University, Changchun 130012, China. He is also with the College of Computing and Data Science, Nanyang Technological University, Singapore 639798 (E-mail: sungeng@jlu.edu.cn). 

  \par Jiacheng Wang and Dusit Niyato are with the College of Computing and Data Science, Nanyang Technological University, Singapore 639798 (E-mails: jiacheng.wang@ntu.edu.sg; dniyato@ntu.edu.sg). 

  \par Shiwen Mao is with the Department of Electrical and Computer Engineering, Auburn University, Auburn, AL 36849-5201 USA (e-mail: smao@ieee.org).
  
  \par Abbas Jamalipour is with the School of Electrical and Computer Engineering, The University of Sydney, Sydney, NSW 2006, Australia (E-mail: a.jamalipour@ieee.org). 

  \par Part of this paper appeared in IEEE MSN 2024~\cite{lai2024irs}.

  \par \textit{(Corresponding author: Geng Sun.)}
  }
}

	

\IEEEtitleabstractindextext{%
\begin{abstract}
Low-altitude wireless networks (LAWNs) represent a promising architecture that integrates unmanned aerial vehicles (UAVs) as aerial nodes to provide enhanced coverage, reliability, and throughput for diverse applications. However, these networks face significant security vulnerabilities from both known and potential unknown eavesdroppers, which may threaten data confidentiality and system integrity. To solve this critical issue, we propose a novel secure communication framework for LAWNs where the selected UAVs within a swarm function as a virtual antenna array (VAA), complemented by intelligent reflecting surface (IRS) to create a robust defense against eavesdropping attacks. Specifically, we formulate a multi-objective optimization problem that simultaneously maximizes the secrecy rate while minimizing the maximum sidelobe level and total energy consumption, requiring joint optimization of UAV excitation current weights, flight trajectories, and IRS phase shifts. This problem presents significant difficulties due to the dynamic nature of the system and heterogeneous components. Thus, we first transform the problem into a heterogeneous Markov decision process (MDP). Then, we propose a heterogeneous multi-agent control approach (HMCA) that integrates a dedicated IRS control policy with a multi-agent soft actor-critic framework for UAV control, which enables coordinated operation across heterogeneous network elements. Simulation results show that the proposed HMCA achieves superior performance compared to baseline approaches in terms of secrecy rate improvement, sidelobe suppression, and energy efficiency. Furthermore, we find that the collaborative and passive beamforming synergy between VAA and IRS creates robust security guarantees when the number of UAVs increases.
\end{abstract}

}

\maketitle
\IEEEdisplaynontitleabstractindextext
\IEEEpeerreviewmaketitle

%
\section{Introduction}\label{sec:introduction}

\par Low-altitude wireless networks (LAWNs) have emerged as a promising paradigm that leverages unmanned aerial vehicles (UAVs) as aerial communication nodes to provide flexible, on-demand connectivity with enhanced coverage and capacity~\cite{yuan2025ground}. Specifically, UAV-based communication schemes in LAWNs are extensively implemented in various key applications of the promising networks, including the Internet-of-things (IoTs), remote sensing, surveillance, agriculture, and connectivity for temporary events. For instance, UAVs function as mobile servers or relays, thus assisting the mobile edge computing systems~\cite{SunTMC}. Moreover, UAVs serve as a new sensing platform, which enables harvesting of various monitoring data~\cite{Long2024}. In addition, UAVs can be equipped with base stations, so that providing network support for the ground users in network-inadequate areas~\cite{Hoang2025TON}. However, the presence of known and potential unknown eavesdroppers poses significant security risks to UAV networks, particularly in LAWNs, where the broadcast nature of wireless communications and the exposure of air-to-ground channels make them vulnerable to interception~\cite{jin2025predictive}.

\par In such cases, physical layer security (PLS) is an effective method that exploits the characteristics of the channel to reduce the efficiency of eavesdroppers in stealing information, and consequently ensures secure communications in LAWNs. Among various PLS methods, collaborative beamforming (CB) is feasible for implementing PLS in UAV communications within LAWN architectures~\cite{Sun2025}. Specifically, by forming a virtual antenna array (VAA), the UAVs in a LAWN can utilize CB to significantly enhance the signal strength for the legitimate receivers, and they can also dynamically adjust the trajectories to optimize the beam pattern of the VAA, thereby effectively suppressing the signal in the direction of potential eavesdroppers. For instance, the authors in~\cite{Sun2022} employed an evolutionary computation-based method to optimize VAAs, which subsequently achieves security and energy efficiency objectives. The authors in~\cite{Zheng2024} proposed a CB-based approach to optimize the communication links in UAV swarms, thereby enhancing the transmission performance of the base station while significantly reducing the corresponding energy consumption and interference to adjacent aerial users. The authors in~\cite{ZhangCb2} investigated the use of UAV swarms to assist secure communication between micro base stations and IoT devices through CB. Moreover, the authors in~\cite{LiCb3} applied CB to both IoT and UAV for realizing the energy and time-efficient data harvesting and dissemination from multiple IoT clusters to remote base stations.

\par However, the CB-based PLS methods may not be suitable in some specific scenarios. For example, if there are buildings or eavesdroppers in the mainlobe direction of VAA, the CB-based PLS methods may become ineffective. In such cases, an intelligent reflecting surface (IRS)~\cite{IRS_wuqingqing, IRS_wuqingqing2} can be a promising solution for these complex environmental circumstances. Specifically, an IRS consists of a multitude of energy-efficient and cost-effective reconfigurable reflecting elements, and by jointly adjusting the phase shifts of these elements, passive beamforming (PB) is achieved, which consequently results in the phase alignment of signals from different transmission paths at the receiver. As such, employing the IRS is beneficial to achieve the PLS in various communication systems. For example, the authors in~\cite{Saif2024} proposed a method to maximize UAV network connectivity under the quality of service constraints, which involves optimizing the deployment of IRS and configuring the virtual partitions in a coordinated manner. Moreover, the authors in~\cite{Li2024a} explored the enhancement of the secure computational performance of wireless power transfer systems in the presence of passive eavesdroppers using IRS, thus maximizing the secure computational task bits for users. In addition, the authors in~\cite{Yi2025} introduced a unified near-far field channel model for IRS-assisted secure networks. They derived a closed-form coverage probability by incorporating IRS physical parameters to provide theoretical and practical co-design guidance for network deployment. Therefore, the UAV arrays can effectively leverage deployed IRS to simultaneously perform CB and PB, thereby enabling PLS while circumventing obstacles.  

\par Nevertheless, implementing such CB and PB systems with UAV swarm-IRS collaborative secure communication system is not straightforward. Specifically, although the mainlobe direction can be shifted away via IRS, the eavesdroppers can still steal overflow signals from the VAA or IRS. Moreover, since UAV is an energy-sensitive system, the energy efficiency of UAV when cooperating with the IRS should be carefully considered. In addition, such systems involve dynamic scenarios, which means that offline optimization methods are not applicable. In particular, eavesdroppers are usually difficult to be detected, uncontrollable, and unpredictable, thereby causing approaches that rely on prior knowledge of the environment to be inadequate for addressing communication security issues. Thus, it is necessary to investigate a novel online optimization approach for adaptively controlling the construction of VAA and the reflection of IRS.

\par Deep reinforcement learning (DRL)-based method can be a promising method for such online and long-term scenarios. For instance, the authors in~\cite{Ning} explored distributed trajectory control for multiple UAVs within a UAV-assisted mobile edge computing (MEC) network. Specifically, they utilized a Nash equilibrium via a game between service providers and implemented a DRL-based algorithm, which consequently minimizes the short-term and long-term computational costs for ground users and UAVs, respectively. Moreover, the authors in~\cite{Hao} considered the task offloading in a UAV-assisted MEC system with multiple collaborating UAVs, and they introduced a novel DRL algorithm based on potential space, thus optimizing the energy consumption and task latency. The authors in~\cite{zhao2022multidrl1} solved the task offloading problem by jointly designing the trajectory of the UAV, computing the task assignment, and managing the communication resources. In addition, the authors in~\cite{Ye} investigated a novel deep recurrent graph network, based on local information about each UAV, which guides a swarm of UAVs over unexplored target areas in partially observable situations. The authors in~\cite{Tariq2024} studied the use of the efficient deep deterministic policy gradient (DDPG) algorithm to determine the optimal trajectory of IRS-assisted UAVs and phase shift angles of IRS, which helps mitigate the disruptive effects of interferers and ensure reliable and secure communications in dynamic environments. Nevertheless, the considered system contains two different types of devices, and this heterogeneity introduces additional challenges to the DRL-based optimization methods.

\par As such, the analyses above motivate us to address the following key issues in the considered joint CB and PB system. \textit{First}, the security and energy efficiency of UAVs are the two most critical goals of this system. Therefore, what approach can be employed to achieve a suitable balance between these competing objectives? \textit{Second}, due to the complex wireless channel and the potential mobility of ground users and eavesdroppers, the system exhibits high dynamics. Hence, how to effectively address the inherently dynamic characteristics of the system? \textit{Finally}, the system contains a large number of decision variables over time (\textit{e.g.,} the real-time trajectories of the UAVs and real-time phase shifts of IRS). These decision variables originate from heterogeneous devices comprising multiple UAVs and IRS. How to optimize these diverse decision variables to ensure the coordination across the system? Accordingly, we consider both the security and energy efficiency of the system and aim to propose a novel DRL-based approach to control the aforementioned decision variables. The main contributions of our work are as follows:

\begin{itemize}

     \item \textit{Energy-efficient UAV Swarm-IRS Collaborative Secure Communication Architecture:} We propose a novel PLS platform that integrates UAV swarm CB with IRS to provide enhanced secure communications for the ground users under eavesdropper threats. Specifically, our architecture dynamically adjusts both CB (\textit{through UAVs}) and PB (\textit{through IRS}) to maximize security while minimizing the energy consumption of UAVs. This architecture is specifically suitable for the highly dynamic environments characterized by mobile users and unpredictable eavesdropper presence. To the best of our knowledge, this is the first security framework that simultaneously leverages the mobility advantages of UAV-based CB with the energy efficiency of passive IRS reflection.

    \item \textit{Joint Collaborative and Passive Beamforming Optimization Problem (JCPBOP):} We model the system to capture the complex interactions between UAV trajectory planning, CB, and IRS phase shift optimization under security constraints. Our analysis reveals critical trade-offs between communication security and energy efficiency. Then, we formulate a JCPBOP to simultaneously maximize the secrecy rate and minimize the maximum sidelobe level (SLL) and total energy consumption by optimizing the excitation current weights and trajectories of the UAVs, as well as the phase shifts of IRS. This problem is a complex NP-hard problem with high dynamics and unpredictability. 
  
    \item \textit{Enhanced Heterogeneous Multi-agent DRL (MADRL)-based Approach:} Due to the highly dynamic and heterogeneous natures of the problem, we first transform the problem into a heterogeneous Markov decision process (MDP). Subsequently, we propose a novel MADRL-based heterogeneous multi-agent control approach (HMCA), which consists of an IRS control policy and an enhanced multi-agent soft actor-critic (MASAC) UAV control policy, thus enabling effective coordination between different types of agents.
  
    \item \textit{Simulations and Analyses:} Simulation results show that the proposed HMCA outperforms various benchmarks in terms of secrecy rate improvement, sidelobe suppression, and energy efficiency. Moreover, the convergence analysis further reveals that HMCA exhibits faster training convergence, which requires approximately 20\% fewer training episodes to reach stable performance compared to baseline methods. In addition, we reveal that the proposed method achieves a more secure performance when employing a higher number of UAVs.
\end{itemize}


\par The rest of this paper is organized as follows. Section \ref{sec:system_model_problem_forlulation} gives the system models and formulates the optimization problem. Section \ref{sec:algorithm} proposes the MADRL-based approach. Section \ref{sec:simulation} shows the simulation results, and Section \ref{sec:conclusion} concludes the overall paper.

%
\section{System Model and Problem Formulation}\label{sec:system_model_problem_forlulation}

%
\begin{figure}
    \centering
    \includegraphics[width=1 \linewidth]{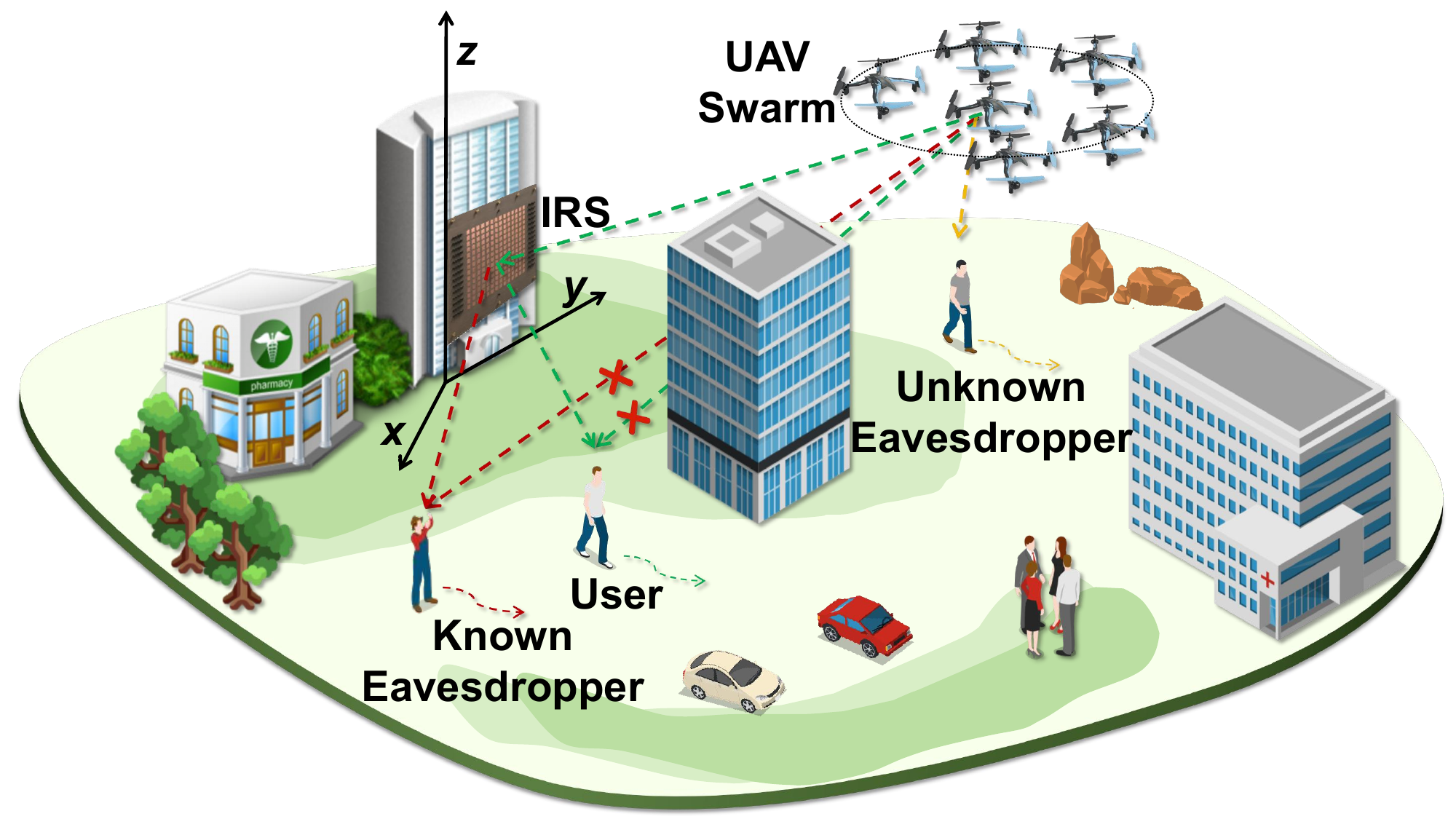}
    \caption{The considered UAV swarm-IRS collaborative secure communication system.}
    \label{Fig:system_model}
\end{figure}

\par In this section, we present the models to characterize the relationships between the decision parameters of the considered system in terms of secure communications or energy consumption. Then, we formulate the JCPBOP.

%
%
\begin{table}[]
\centering
\renewcommand\arraystretch{1.15}
\caption{Notation}
\label{table:notation}
\resizebox{3.5in}{!}{
\begin{tabular}{@{}ll@{}}
\toprule
\textbf{Notation}           & \textbf{Description}                                 \\ \midrule
$\alpha$                    & Temperature parameter                                \\
$\alpha_d$                  & Path loss exponent of the U-G direct link            \\
$\alpha_r$                  & Path loss exponent of the R-G link                   \\
$\beta$                     & Rician factor of the R-G link                        \\
$\gamma$                    & Discount factor                                      \\
$\theta_m$                  & Actor network parameters of agent $m$                \\
$\lambda$                   & Carrier wavelength                                   \\
$\varrho_t^{\mathrm{AR}}$   & Sine value of horizontal AoA from VAA to IRS         \\
$\varrho_t^{\mathrm{RU}}$   & Sine value of horizontal AoA from IRS to user        \\
$\tau$                      & Target network updating rate                         \\
$\varphi_t^{{\mathrm{AR}}}$ & Sine value of vertical AoA from VAA to IRS           \\
$\varphi_t^{{\mathrm{RU}}}$ & Sine value of vertical AoD from IRS to user          \\
$\chi_t^{{\mathrm{AR}}}$    & Cosine value of horizontal AoA from VAA to IRS       \\
$\chi_t^{{\mathrm{RU}}}$    & Cosine value of horizontal AoA from IRS to user      \\
$\psi_m$                    & Critic network parameters of agent $m$               \\
$\bm{a}_t^m$                & Action of agent $m$                                  \\
$b_t^m$                     & Incentive reward for agent $m$                       \\
$\mathcal{B}$               & Replay buffer                                        \\
$N_M$                       & Number of UAVs                                       \\
$N_S$                       & Total number of reflecting elements of IRS       \\
$N_{S}^\mathrm{C}$          & Number of reflecting elements in a column of IRS     \\
$N_{S}^\mathrm{R}$          & Number of reflecting elements in a row of IRS        \\
$Q_{\psi_m}(\bm s,\bm a)$   & Q-function with ${\psi}_m$ for agent $m$             \\
$r_t^m$                     & Reward for agent $m$                                 \\
$s_c$                       & Column index of the $s$th reflecting elements in IRS \\
$s_r$                       & Row index of the $s$th reflecting elements in IRS    \\
$w_{s}$                     & Phase shift control of the $s$th reflecting elements \\ \bottomrule
\end{tabular}
}
\end{table}

%
\subsection{System Overview}

\par As illustrated in Fig. \ref{Fig:system_model}, we consider a UAV swarm-IRS collaborative secure communication system. Specifically, a set of UAVs denoted as $\mathcal{M}\triangleq \{1,\ldots, m,\ldots, N_M\}$ provides downlink communications to a ground user denoted as $U$. However, several eavesdroppers, denoted by $\mathcal{V}$, are present between the UAVs and the ground user, who attempt to intercept the signals and thereby pose a threat to the communication. Moreover, an IRS which is equipped with a set of passive reflecting elements denoted as $\mathcal{S}\triangleq \{1, \ldots,s,\ldots,N_S\}$ is deployed to assist the communication process. We consider that the UAVs and IRS can be controlled by a high-performance controller (\textit{e.g.}, field programmable gate arrays). In addition, in the considered system, the ground user and eavesdroppers are movable to meet the actual situation. Thus, we employ the Gauss-Markov random movement model~\cite{yang2022onlineGMRMM} to simulate the motion trajectories of the user and the eavesdropper.

\par Due to the existence of object occlusion and eavesdroppers, the UAVs will construct a VAA and subsequently direct the signals to the IRS. Following this, the IRS will reflect the signals to the ground user via PB, while during this process, the eavesdropper can monitor the links in both VAA-to-IRS and IRS-to-user channels. 

\par Without loss of generality, we consider a three-dimensional (3D) Cartesian coordinate. We also consider a discrete-time system that evolves over timeslot $\mathcal{T}=\{1,..., t,..., T\}$. As such, the coordinate of the $m$th UAV is represented as ${(x_m(t),y_m(t),z_m(t))}$. Moreover, the location of IRS is not time-varying and is therefore denoted as $(x_\mathrm{R},y_\mathrm{R},z_\mathrm{R})$. 

\par In the following sections, we will first introduce VAA and IRS models to identify the decision variables, and then subsequently introduce transmission and energy models to derive the main objectives.

%
%
\subsection{VAA Model}
\label{ssec:vaa_model}

\par In a VAA, each UAV is treated as an individual antenna element within the array to implement CB. The electromagnetic waves transmitted by these UAVs superimpose or cancel each other, which results in a high gain in specific directions. Mathematically, the array factor is employed to analyze the signal distribution. Let $I_m$ denote the excitation current weight of the $m$th UAV, and then the array factor of VAA is thus given by~\cite{Li2024tmc}
\begin{equation}
\label{eq:af}
    AF(\theta, \phi)=\sum_{m=1}^{N_M} I_m e^{j\left[k_c\left(x_m \sin \theta \cos \phi+y_m \sin \theta \sin \phi+z_m \cos \theta\right)\right]},
\end{equation}

\noindent where $\theta \in [ 0,\pi ]$ and $\phi \in[-\pi,\pi ]$ are the horizontal and vertical angles, respectively. Additionally, $ k_c = 2\pi/\lambda $ is the phase constant, where $\lambda$ is the wavelength. The array factor is directly affected by the locations and excitation current weights of the UAVs~\cite{Sun2021}. Furthermore, the UAVs in the VAA utilize the methods in~\cite{alemdar2021rfclockAF}~\cite{Mohanti2022} to perform frequency, phase, and time synchronizations.

\par Then, we derive the antenna gain via array factor for further measuring the transmission performance of VAA. Let $\theta_{rec}$ and $\phi_{rec}$ represent the horizontal and vertical angles of the receiver, respectively. As such, the transmission gain of VAA for CB is therefore expressed as follows~\cite{Li2023TON}:
\begin{equation}
    \begin{aligned}
        G_{rec}=
        \frac{4 \pi\left|AF\left(\theta_{rec}, \phi_{rec}\right)\right|^2 w\left(\theta_{r e c}, \phi_{r e c}\right)^2}{\int_0^{2 \pi} \int_0^\pi|AF(\theta, \phi)|^2 w(\theta, \phi)^2 \sin \theta \mathrm{d} \theta \mathrm{d} \phi} \eta,
        \label{eq:gain}
    \end{aligned}
\end{equation}
\noindent where $ w(\theta,\phi) $ is the far-field beam pattern. In addition, $\eta \in [0,1] $ represents the antenna array efficiency. 

\par From Eq.~\eqref{eq:gain}, it is observed that the array factor of VAA determines the transmission gain. Moreover, Eq.~\eqref{eq:af} indicates that the excitation current weights and coordinates of the UAVs influence the array factor, thereby establishing them as critical parameters for optimizing secure performance of the CB transmission.

%
%
\subsection{IRS Model}
\label{ssec:Passive Beamforming}

\par An IRS is considered that consists of $N_{S}^\mathrm{R} \times N_{S}^\mathrm{C}$ passive reflecting elements with continuously controllable phase shifts, which are arranged as a uniform planar array to enable PB. Let $w_s(t) \in [0,2\pi)$ represent the phase shift of the $s$th element in timeslot $t$, and then the diagonal matrix of IRS phase shifts for PB is denoted as follows:
\begin{equation}
    \bm{\Theta}{(t)} \triangleq \mathrm{diag}(e^{jw_1(t)},e^{jw_2(t)},\ldots,e^{jw_{N_S}(t)}) \in \mathbb{C}^{N_S\times N_S}.
\end{equation}

\par In addition to the number of reflecting elements and their respective phase shifts, the geometry of IRS and the configuration of the reflecting elements also significantly influence the characteristics of reflection for PB. For this model, the reflection amplitude of each reflecting element is set as the maximum value, thereby achieving the maximum user receiving power\cite{IRS_wuqingqing2}. With this configuration, the optimal IRS phase shifts $\bm{\Theta}{(t)}$ can depend on the cascaded channels via the IRS~\cite{IRS_wuqingqing2}.

\subsection{Channel Model} 
\label{ssec:channel_model}

\par The considered system consists of multiple communication links with distinct channel fading characteristics. Specifically, the system incorporates the channels from UAV to IRS, from IRS to the user, and from UAV to the user. Furthermore, the channels from UAV to the eavesdropper and from IRS to the eavesdropper are taken into account. Since these channels exhibit different physical characteristics, each channel will be presented separately in the following.

%
%
\subsubsection{UAV-to-IRS Channel} 

\par Due to the use of IRS to avoid obstacles and the high altitude of UAVs, a line-of-sight (LoS) channel between VAA and IRS is considered. Let $\phi_{\mathrm R}(t)$ and $\theta_{\mathrm R}(t)$ denote the horizontal and vertical angles of departure (AoD) from the VAA to the IRS, then the channel gain from the VAA to the IRS is~\cite{IRS_wuqingqing}
\begin{equation}
    \begin{aligned}
        \bm{h}&_{\mathrm{AR}}{(t)} =  AF(\phi_{\mathrm R}(t),\theta_{\mathrm R}(t)) \sqrt{\rho d_{\mathrm{AR}}^{-2}{(t)}} \bm{h}_{\mathrm{AR}}^{\mathrm{LoS}}{(t)},
        \label{Eq_channel_vaa_irs}
    \end{aligned}
\end{equation}
\noindent where $\rho$ denotes the path loss coefficient and $d_{\mathrm{AR}}(t)$ denotes the distance between the VAA center and the IRS at timeslot $t$. Another key component of Eq.~\eqref{Eq_channel_vaa_irs} is the LoS component $\bm{h}_{\mathrm{AR}}^{\mathrm{LoS}}{(t)}$. Specifically,  $\bm{h}_{\mathrm{AR}}^{\mathrm{LoS}}{(t)} \in \mathbb{C}^{N_S\times 1}$ is given by
\begin{equation}
    \begin{aligned}
        \bm{h}_{\mathrm{AR}}^{\mathrm{LoS}}{(t)} \!&=\!  {[1, e^{-j\frac{2 \pi d_r}{\lambda_c}}\chi_t^{\mathrm{AR}}\varphi_t^{\mathrm{AR}}, \ldots, e^{-j \frac{2 \pi (N_{S}^\mathrm{R}-1) d_{\mathrm{r}}}{\lambda_c}  }\chi_t^{\mathrm{AR}}\varphi_t^{\mathrm{AR}}]^{\mathrm{T}} } \\
         &\otimes[1, e^{-j \frac{2 \pi d_c}{\lambda_c}} \varrho _t^{\mathrm{AR}}\varphi_t^{\mathrm{AR}}, \ldots, e^{-j \frac{2 \pi(N_{S}^\mathrm{C}-1) d_{\mathrm{c}}}{{\lambda_c} } }\varrho _t^{\mathrm{AR}}\varphi_t^{\mathrm{AR}}]^{\mathrm{T}},
    \end{aligned}
    \label{eq:vaa_irs_los}
\end{equation}
\noindent where $\lambda_c$, $d_r$, and $d_c$ denote the carrier wavelength, row separation, and column separation of IRS reflecting elements, respectively. Moreover, $\varphi_t^{{\mathrm{AR}}}$, $\chi_t^{{\mathrm{AR}}}$, and $\varrho _t^{\mathrm{\mathrm{AR}}}$ represent the sine value of vertical angle of arrival (AoA), the cosine and sine values of the horizontal AoAs of the signal from the VAA to IRS in timeslot $t$, respectively. Note that the calculations of these three parameters follow~\cite{IRS_sum_simple_method}.

%
%
\subsubsection{IRS-to-User Channel} 

\par According to~\cite{li2020reconfigurableIRS-classic_los-channel}, the channel from the IRS to the user is modeled by a Rician fading channel, which is given by~\cite{HanUAVIRSSecrecy}
\begin{equation}
    \begin{aligned}
        \bm{h}&_{\mathrm{RU}}{(t)} = \\& \sqrt{\rho d_{\mathrm{RU}}^{-\alpha_r}{(t)}} \left(\sqrt{\frac{\beta}{1+\beta}} \bm{h}_{\mathrm{RU}}^{\mathrm{LoS}}{(t)}+\sqrt{\frac{1}{1+\beta}} \bm{h}_{\mathrm{RU}}^{\mathrm{NLoS}}{(t)}\right),
    \end{aligned}
    \label{Eq_channel_irs_user}
\end{equation}
\noindent where $d_{\mathrm{RU}}(t)$, $\alpha_r$, $\bm{h}_{\mathrm{RU}}^{\mathrm{LoS}}{(t)}$ and $\bm{h}_{\mathrm{RU}}^{\mathrm{NLoS}}{(t)}$ denote the distance between the user and the IRS, the corresponding path loss exponent, the LoS and Non-LoS components, respectively. Specifically, $\bm{h}_{\mathrm{RU}}^{\mathrm{LoS}}{(t)} \in \mathbb{C}^{N_S\times 1}$ can be given by
\begin{equation}
    \begin{aligned}
        \bm{h}_{\mathrm{RU}}^{\mathrm{LoS}}{(t)} \!&=\! {[1, e^{-j\frac{2 \pi s_r}{\lambda_c}}\chi_t^\mathrm{RU}\varphi_t^\mathrm{RU}, \ldots, e^{-j \frac{2 \pi (N_{S}^{R}-1) d_{\mathrm{r}}}{\lambda_c}}\chi_t^\mathrm{RU}\varphi_t^\mathrm{RU}]^{\mathrm{T}} } \\
        &\otimes[1, e^{-j \frac{2 \pi d_c}{\lambda_c}} \varrho _t^\mathrm{RU}\varphi_t^\mathrm{RU}, \ldots, e^{-j \frac{2 \pi (N_{S}^{C}-1) d_{\mathrm{c}}}{{\lambda_c}}}\varrho_t^\mathrm{RU}\varphi_t^\mathrm{RU}]^{\mathrm{T}},
    \end{aligned}
    \label{eq:irs_user_los}
\end{equation}

\noindent where $\varphi_t^{\mathrm{\mathrm{RU}}}$, $\chi_t^{\mathrm{\mathrm{RU}}}$ and $\varrho_t^{\mathrm{\mathrm{RU}}}$ represent the sine value of vertical AoD, the cosine and sine values of the horizontal AoDs of the signal from IRS to the ground user in timeslot $t$, respectively. Additionally, the non-LoS component $\bm{h}_{\mathrm{RU}}^{\mathrm{NLoS}}{(t)}$ of Eq.~\eqref{Eq_channel_irs_user} is characterized by independent samples generated from a complex Gaussian distribution that exhibits circular symmetry with zero mean and unit variance.

%
%
\subsubsection{UAV-to-User Channel} 

\par According to~\cite{li2020reconfigurableIRS-classic_los-channel}, the channel from the VAA to the user can be expressed as a Rayleigh fading model. The channel gain of the VAA-to-user link can be expressed as
\begin{equation}
    h_{\mathrm{AU}}{(t)} = AF(\phi_{u}(t),\theta_{u}(t))\sqrt{\rho d_{\mathrm{AU}}^{-\alpha_d}{(t)}}\tilde h,
    \label{eq:vaa_user_channel}
\end{equation}

\noindent where $d_{\mathrm{AU}}{(t)}$, $\alpha_d$, $\phi_{u}(t)$ and $\theta_{u}(t)$ denote the distance between the VAA and the user, the corresponding path loss exponent, the horizontal and vertical AoDs from the VAA to the user, respectively. Additionally, the term $ \tilde h $ is represented mathematically as a complex Gaussian random variable that maintains circular symmetry properties.

\subsection{Secure Transmission Models}
\label{ssec:transmission_model}

\par Based on the aforementioned VAA, IRS, and channel models, we present the transmission model and define the secure communication metric. 
 
\par \textit{First}, we define the secrecy rate to evaluate the secure performance of the communications under the threat of the known eavesdropper~\cite{wang2025graph}. Specifically, by integrating the transmission gain of VAA, the channel power gains, and the PB gain of IRS, we can articulate the total gain of the communications as follows~\cite{IRS_wuqingqing}~\cite{Li2023TON}:
\begin{equation}
    \begin{aligned}
        G_U(t)  = \frac{4\pi \left| (\bm{h}_{\mathrm{AR}}{(t)})^T\bm{\Theta}{(t)}\bm{h}_{\mathrm{RU}}{(t)}  +h_{\mathrm{AU}}{(t)} \right|^2}
         {{\int_0^{2 \pi} \int_0^\pi|AF(\theta, \phi)|^2 w(\theta, \phi)^2 \sin \theta \mathrm{d} \theta \mathrm{d} \phi}}\eta.
    \end{aligned}
    \label{eq:gain2}
\end{equation}

\par Subsequently, the achievable rate of the user can be defined as follows:
\begin{equation}
    R_{U}{(t)} = B\log_2(1+\frac{P_t G_U(t)}{\sigma^2}),
    \label{eq:transmission_rate}
\end{equation}
\noindent where $B$, $P_t$, and $\sigma^2$ denote the transmission bandwidth, total transmit power of VAA, and noise power, respectively. 

\par Similarly, for the eavesdropper, the channel gain $G_E(t)$ can be calculated by replacing the user-related channel vectors in Eq.~\eqref{eq:gain2} with the corresponding eavesdropper channel vectors:
\begin{equation}
    \begin{aligned}
        G_E(t)  = \frac{4\pi \left| (\bm{h}_{\mathrm{AR}}{(t)})^T\bm{\Theta}{(t)}\bm{h}_{\mathrm{RE}}{(t)}  +h_{\mathrm{AE}}{(t)} \right|^2}
         {{\int_0^{2 \pi} \int_0^\pi|AF(\theta, \phi)|^2 w(\theta, \phi)^2 \sin \theta \mathrm{d} \theta \mathrm{d} \phi}}\eta.
    \end{aligned}
    \label{eq:gain_e}
\end{equation}

\par The achievable rate of the eavesdropper can then be calculated using the same approach as in Eq.~\eqref{eq:transmission_rate}:
\begin{equation}
    R_{E}{(t)} = B\log_2(1+\frac{P_t G_E(t)}{\sigma^2}).
    \label{eq:transmission_rate_e}
\end{equation}

\par Consequently, we define the secrecy rate as the difference between the achievable rate of the legitimate user and that of the eavesdropper:
\begin{equation}
    R_{sec}{(t)} =  R_{U}{(t)} - R_{E}{(t)}.
    \label{eq:obj1}
\end{equation}

\par \textit{Second}, we adopt the maximum SLL to reflect the threat of the unknown eavesdropper. Specifically, the maximum SLL of VAA in timeslot $t$ can be expressed as follows:
\begin{equation}
    \begin{aligned}
        R_{SL}(t) = \frac{\max \left| AF(\theta_{SL}{(t)},\phi_{SL}
            (t))\right| }{AF(\theta_{ML}{(t)},\phi_{ML}{(t)}) },
    \end{aligned}
    \label{eq:obj2}
\end{equation}
\noindent where $(\theta_{SL}{(t)},\phi_{SL}{(t)})$ and $(\theta_{ML}{(t)},\phi_{ML}{(t)})$ are the total SLL directions and mainlobe direction, respectively. 

\par From Eqs.~\eqref{eq:gain2}, \eqref{eq:gain_e}, and~\eqref{eq:obj2}, we can observe that $\bm{\Theta}(t)$ is a key parameter that affects both the user gain and the eavesdropper gain, and thus influences the secrecy rate. Therefore, optimizing $\bm{\Theta}(t)$ is essential for enhancing the secure performance of the considered system.

\subsection{UAV Movable Model} 
\label{ssec:energy_model}

\par During the construction of VAA, the UAVs are required to fine-tune their positions. Consequently, we present the movement and energy model of the UAV as follows.

\par In timeslot $t$, we consider the $m$th UAV can fly towards a horizontal flight direction $\phi_m(t) \in [0,2\pi)$ with the horizontal flight speed $v_m(t) \in [0,v_{max}]$, and $v_{max}$ is the maximum horizontal speed of the UAV. Moreover, we also consider the arbitrary 3D trajectory with the UAV climbing and descending over time in vertical speed $l_m(t) \in [-l_{max},l_{max}]$, and $l_{max}$ is the maximum climbing or descending speed of the UAV. The mobility of a UAV in timeslot $t$ is expressed as
\begin{equation}
    \label{eq:UAV_mobility}
    \left\{\begin{array}{l}
x_{m}(t)=x_{m}(0)+\sum_{t^{\prime}=1}^{t} v_{m}\left(t^{\prime}\right) \cos \left(\phi_{m}\left(t^{\prime}\right)\right) \\
y_{m}(t)=y_{m}(0)+\sum_{t^{\prime}=1}^{t} v_{m}\left(t^{\prime}\right) \sin \left(\phi_{m}\left(t^{\prime}\right)\right)\\
z_{m}(t)=z_{m}(0)+\sum_{t^{\prime}=1}^{t} l_{m}\left(t^{\prime}\right)
\end{array}\right. .
\end{equation}

\par Moreover, the movement distance of the UAV in timeslot $t$ can be expressed as follows:
\begin{equation}
    \label{eq:uav_fly_distance}
    d_m(t) = \sqrt{{\triangle x_m(t)}^2 + {\triangle y_m(t)}^2 + {\triangle z_m(t)}^2},
\end{equation}
\noindent where $\triangle x_m(t) $, $\triangle y_m(t)$, $\triangle z_m(t)$ represent the 3D coordinates difference of the UAV in timeslot $t$, respectively.

\par Furthermore, let $\triangle t$ denote the duration of each timeslot, and the average speed of the UAV flying in 3D space of timeslot $t$ can be expressed as
\begin{equation}
    \label{eq:uav_average_speed}
    \bar v_{m}(t) = \frac{d_m(t)}{\triangle t}.
\end{equation}

\par In case of a rotary-wing UAV flying in two-dimensional horizontal space, the energy consumption for propulsion and overcoming gravity is much greater than the communication energy consumption. Moreover, the UAV propulsion power is related to its speed, \textit{i.e.}~\cite{Ma2024},
\begin{equation}
    \begin{aligned}
        P(\bar v_m)=& P_B\left(1+\frac{3 \bar v_m^2}{v_{t i p}^2}\right)+P_I\left(\sqrt{1+\frac{\bar v_m^4}{4 v_0^4}}-\frac{\bar v_m^2}{2 v_0^2}\right)^{1 / 2}+ \\
        & \frac{1}{2} d_0 \rho s A \bar v_m^3,
    \end{aligned}
    \label{eq:uav_power}
\end{equation}
\noindent where the parameters $P_B$ and $P_I$ represent two distinct constant values associated with the blade profile power and induced power during hover operations, respectively. The symbol $v_{t i p}$ denotes the peripheral velocity at the rotor blade's extremity, while $v_0$ signifies the average velocity induced by the rotor when the UAV maintains a stationary hovering position. The coefficients $d_0$ and $s$ correspond to the aircraft's fuselage drag ratio and the rotor's solidity parameter, respectively. The physical environmental parameter $\rho$ characterizes the atmospheric density, and $A$ represents the total surface area of the rotor disc.

\par Besides, the energy consumption of the UAV in timeslot $t$ in 3D flight is expressed as follows:
\begin{equation}
    \begin{aligned}
        E_m(t) \approx & P(v_m(t))\triangle t+\frac{1}{2} m_{M}\left(\bar v_m(t)^2-\bar v_m(t-1)^2\right)+\\
        & m_{M} g(z_m(t)-z_m(t-1)),
    \end{aligned}
    \label{eq:uav_energy}
\end{equation}
\noindent where the parameter $m_M$ represents the total physical mass of the unmanned aerial vehicle, and $g$ denotes the gravitational acceleration.

%

\subsection{Problem Formulation}

\par The main objective is to improve the secure performance of the considered system while minimizing the energy consumption of UAV swarm. 

\par Note that there exist trade-offs between communication security and energy efficiency in the considered system. Specifically, enhancing the secrecy performance typically requires sophisticated UAV trajectory designs and CB strategies, which consequently increase energy consumption. Conversely, minimizing energy consumption might constrain UAVs to less optimal positions, thereby potentially compromising the secrecy performance of the system. Therefore, finding a reasonable balance between these conflicting objectives becomes paramount for the considered UAV swarm-IRS collaborative secure communication. 

\par To this end, we need to carefully optimize the key decision variables related to secure performance and energy efficiency. Specifically, the secrecy rate and maximum SLL of the considered system are jointly determined by the trajectories and excitation current weights of UAVs, IRS phase shifts, and channel power gains over time. Furthermore, the energy efficiency of the UAV swarm is intrinsically linked to their trajectories throughout the task. 

\par As such, we let $\bm{I} = \{I_m(t) \mid m \in \mathcal{M}\} $, $ \bm{v} = \{v_m(t) \mid m \in \mathcal{M}\}$, $ \bm{\phi} = \{\phi_m(t) \mid m \in \mathcal{M}\}$ and $\bm{l} =\{l_m(t) \mid m \in \mathcal{M}\}$ denote the set of excitation current weights, horizontal speed set, horizontal flight direction and vertical speed of all UAVs, respectively. Then, the optimization objectives are as follows.

\par \textit{First}, we need to maximize the secrecy rate of the system, which ensures secure communication between the UAV swarm and legitimate receivers in the presence of eavesdroppers. Thus, the first optimization objective can be given by
\begin{equation}
  f_1(\bm{I}, \bm{v}, \bm{\phi}, \bm{l}, \bm{\Theta}) = \sum\nolimits_t^{T} R_{sec}(t).
\end{equation}

\par \textit{Then}, we minimize the maximum SLL of VAA, which will help suppress signal leakage in unintended directions, particularly toward potential eavesdroppers. Thus, the second optimization objective is represented as 
\begin{equation}
  f_2(\bm{I}, \bm{v}, \bm{\phi}, \bm{l}, \bm{\Theta}) = \sum\nolimits_t^{T} R_{SL}(t).
\end{equation}

\par \textit{Finally}, we minimize the total energy consumption of the UAV swarm, which is crucial for extending the operational duration of the LAWN. Thus, the third optimziation objective function is designed as
\begin{equation}
  f_3(\bm{I}, \bm{v}, \bm{\phi}, \bm{l}, \bm{\Theta}) = \sum\nolimits_{m=1}^{N_M}{E_m(t)}.
\end{equation}

\par Based on these conflicting objectives, we formulate a JCPBOP, which is a multi-objective optimization problem, as follows:
\begin{subequations}
    \begin{align}
        \max_{\bm{I},\bm{v}, \bm{\phi}, \bm{l}, \bm{\Theta}} &\ F = \{f_1, -f_2, -f_3\},\\ 
        \mathrm{s.t.}\quad
        &0 \le I_m{(t)} \le 1,\forall m \in \mathcal{M}, \label{conji}\\ 
        &L_{min} \le x_m{(t)} \le L_{max},\forall m \in \mathcal{M}, \label{conjx} \\
        &L_{min} \le y_m{(t)} \le L_{max},\forall m \in \mathcal{M}, \label{conjy}\\ 
        &H_{min} \le z_m{(t)} \le H_{max},\forall m \in \mathcal{M}, \label{conjz}\\
        &0 \le w_s(t) < 2\pi,\forall s \in \mathcal{S},  \label{irs_conj1} \\
        &D_{(m_1,m_2)}\geq D_{min}, \forall m_1, m_2 \in \mathcal {M}, \label{collision}
    \end{align}
    \label{problem}
\label{formulation}
\end{subequations}

\vspace{-3 mm}
\noindent where $ L_{min} $, $ L_{max} $, $ H_{min} $, and $ H_{max} $ are the constraints on the UAV movement. Specifically, $ L_{min} $ and $ L_{max} $ define the smallest and largest allowed areas for UAV movement in the horizontal plane, while $ H_{min} $ and $ H_{max} $ set the lowest and highest flying heights in vertical directions. The constraint in \eqref{irs_conj1} controls the phase shift and magnitude settings for IRS reflecting elements. Finally, the constraint in \eqref{collision} requires that any two UAVs must stay at least $D_{min}$ distance apart to avoid collisions during operation.

\par Note that the formualted JCPBOP is challenging for common optimization methods due to the following characteristics:

\begin{itemize}
    \item \textit{High dynamic:} This problem involves dynamic user and eavesdropper motions, multiple varying channels, and random initial UAV locations. Therefore, the algorithm needs to promptly adapt to changing deployment conditions. However, most optimization methods, such as convex or non-convex optimization, require prior information about the environment to make decisions, and thus, they are unsuitable for this problem. 

    \item \textit{Heterogeneity:} The decision variables encompass optimization of UAV swarm and IRS (\textit{e.g.}, as shown in Eqs.~\eqref{eq:af} and~\eqref{eq:gain2}). These two systems have different decision varibles but with mutual dependencies, which further complicates the optimization. 

    \item \textit{Long-term Cumulative Objective Orientation:} This problem seeks to consider the cumulative objective within a time period. However, most of the existing optimization methods cannot effectively balance the short-term and long-term objective of the problem. 
\end{itemize}

\par Given these challenges, MADRL can be a suitable framework for solving the problem. Specifically, MADRL considers each self-organized system of the environment (\textit{e.g.}, UAV or IRS) as an agent. These agents learn to operate in uncertain environments by executing specific actions and subsequently observing the resulting rewards. Through this learning process, they can make sequential decisions that aim at maximizing cumulative rewards. Consequently, MADRL offers significant potential for the real-time deployment of UAV swarm and IRS in dynamic environments. In particular, MADRL can handle both complex actions and can be deployed in a distributed manner due to the introduced neural networks and multi-agents. Thus, in the following sections, we propose an MADRL-based method to solve the problem.

\section{MADRL-based Heterogeneous Control Approach} \label{sec:algorithm}

\par In this section, we first reformulate the formulated problem as a heterogeneous MDP. Then, we introduce the policy for controlling the IRS agent. After that, we present an MADRL-based algorithm for coordinating the UAV swarm. Finally, we propose the framework of the considered HMCA.

%
\subsection{Heterogeneous MDP Formulation}

\par The considered system involves two types of agents, \textit{i.e.}, UAVs and IRS, which have different observations and actions. As such, we transform JCPBOP shown in Eq.~\eqref{problem} as a heterogeneous MDP. This MDP can be denoted as a tuple $(\mathcal{S}, \mathcal{A}, \mathcal{P}, \mathcal{R}, \gamma)$, where each element represents a fundamental aspect of the decision process. Specifically, $\mathcal{S}$ captures the possible system states, $ \mathcal{A} $ encompasses all available actions, $\mathcal{P} $ models the environmental dynamics through transition probabilities, $\mathcal{R} $ quantifies the obtained benefits via the reward mechanism, and $\gamma$ balances immediate versus future rewards. The key components are detailed as follows.

\subsubsection{State Space} 

\par In practice, CB requires data sharing and synchronization among UAV swarm, which means that a UAV can observe the position of other UAVs via the method in~\cite{alemdar2021rfclockAF, Mohanti2022}. Moreover, the locations of the user and eavesdropper are also known to UAVs. Likewise, the UAV swarm will share the environmental information with the IRS. As such, for both UAV and IRS agents, the state space can be defined as follows:
\begin{equation}
    \begin{aligned}
        \bm s_{t} = &\{x_m(t), y_m(t), z_m(t) \mid m \in \mathcal{M}\} \\ & \cup \{x_u(t),y_u(t),x_e(t),y_e(t)\}.
    \end{aligned}
    \label{eq:state}
\end{equation}

\subsubsection{Action Space}

\par Each UAV agent will take actions for adjusting the excitation current weight, horizontal speed, horizontal flight direction, and vertical speed, which is given by $\bm a_t^m = \{{I}_m(t), v_m(t), \phi_m(t), l_m(t)\}$. As such, the action space for all the UAVs is given as follows:
\begin{equation}
    \begin{aligned}
    \bm{a}_t= \left\{\bm a_t^m \mid m \in \mathcal{M} \right\}.
    \label{eq:UAV_action}
    \end{aligned}
\end{equation}

\par Moreover, the IRS agent will perform phase shifts for signal reflection, and the action is given as follows:
\begin{equation}
    \bm{a}_t^\mathrm{R} = \{e^{jw_1(t)},\ldots,e^{jw_s(t)},\ldots,e^{jw_{N_S}(t)}\}.
\end{equation}

\subsubsection{Reward Function} 

\par Reward function is crucial for the effectiveness of MADRL as a control policy for our formulated problem. To achieve optimal performance and guide the behavior of both UAV and IRS agents, we design a comprehensive reward function that incorporates the optimization objectives while addressing the constraints and facilitating efficient agent exploration. As such, the reward function for agent $m$ at time $t$ is given by
\begin{equation}
    \begin{aligned}
    r_{t}^{m}=\left\{\begin{array}{ll}
    \Psi_{m}(t) + b_t^m, & \text {satisfy \eqref{conji}-\eqref{collision}} \\
    -\eta_{1} Z_{m}^{1}- \eta_{2} \sum_{m^{\prime} \in \mathcal{M}} Z_{m m^{\prime}}^{2}, & \text {otherwise}
    \end{array}\right.
    \end{aligned},
    \label{eq:final_reward}
\end{equation}

\noindent where $\Psi_m(t) = \varepsilon_1 R_{sec}(t)+ \varepsilon_2 R_{SL}(t)+ \varepsilon_3 \sum\nolimits_{m=1}^{N_M}{E_m(t)}$ represents the optimization objectives with $\varepsilon_1$, $\varepsilon_2$, and $\varepsilon_3$ being weights indicating the different significance of each objective. 

Moreover, the penalty terms include $\eta_1$ and $\eta_2$ as constants used to penalize UAVs that violate the boundary constraints or collide with other UAVs, respectively. Likewise, the binary variables $Z_m^1$ and $Z_{mm^\prime}^2$ indicate whether UAV $m$ is outside the boundary and whether it collides with UAV $m^\prime$, respectively.

\par In addition, the term $b_t^m$ represents an incentive mechanism designed to enhance exploration efficiency by guiding UAVs toward the IRS while maintaining a safe distance from the flyable boundary:
\begin{equation}
    \begin{aligned}
        b_t^m =  \zeta_1\underbrace{\frac{\bm{w}_t^{m,\mathrm{R}}\cdot {\bm {d}_t^m}} {\Vert{\bm{w}_t^{m,\mathrm{R}}}\Vert \Vert {\bm {d}_t^m} \Vert}}_{\text{Directional term}} -\zeta_2 \underbrace{\Vert{\bm{w}_t^{m,r}\Vert}}_{\text{Positional term}},
    \end{aligned}
\label{eq:incentive_component}
\end{equation}

\noindent where $\bm{w}_t^{m,\mathrm{R}}$ denotes the vector from UAV to the IRS, $\bm {d}_t^m$ represents the displacement of UAV in timeslot $t$, and $\bm{w}_t^{m,r}$ is the vector from the UAV to a reference point $\bm r$ within the feasible flight space. The parameters $\zeta_1$ and $\zeta_2$ are the directional and positional weights for the incentive rewards, respectively.

\par Note that the directional term encourages UAVs to fly toward the IRS, thereby improving transmission performance as demonstrated in \cite{IRS_wuqingqing2}, while the positional term ensures that UAVs maintain a sufficient distance from the flyable boundary, thus enhancing operational safety and exploration efficiency. By carefully designing the reward function, both UAV and IRS agents can efficiently coordinate their actions to optimize the secrecy rate, sidelobe suppression, and energy efficiency of the system.

%
\subsection{Policy for Controlling IRS Agent}
\label{subsection:irs_control_policy}

\par The IRS agent has the following characteristics when responding to environmental changes. \textit{First}, the IRS agent needs to control the phase shifts of a large number of reflecting elements, which means that its action space is large. \textit{Second}, the phase shifts of IRS in different timeslots are relatively independent, and therefore there is no dependence on the previous state. \textit{Finally}, there are some existing works proposing optimal or near-optimal control of the reflecting elements of IRS that depend on the positions of the transmitter and receiver. 

\par Based on the aforementioned reasons, we transform the multiple-timeslot IRS controlling into a single-slot optimization process. In this case, the IRS agent directly utilizes the existing optimization method to calculate the IRS phase shifts based on the locations of the transmitter and receiver of the current timeslot. The main steps are as follows.

\par Let $\varphi_t^{{\mathrm{AR}}}$, $\chi_t^{{\mathrm{AR}}}$, and $\varrho _t^{\mathrm{\mathrm{AR}}}$ represent the sine value of vertical AoA, the cosine and sine values of the horizontal AoAs of the signal from the VAA to IRS in timeslot $t$, respectively. Additionally, let $\varphi_t^{\mathrm{\mathrm{RU}}}$, $\chi_t^{\mathrm{\mathrm{RU}}}$ and $ \varrho _t^{\mathrm{\mathrm{RU}}}$ represent the sine value of vertical AoD, the cosine and sine values of the horizontal AoDs of the signal from IRS to the ground user in timeslot $t$, respectively. As a result, the phase shift of the $s$th reflecting elements $w_{s}(t)$ can be determined as follows~\cite{IRS_sum_simple_method}:
\begin{equation}
\label{eq:irs_op_method}
    \begin{aligned}
        w_{s}(t) = \frac{2 \pi }{\lambda_c} s_{r} d_{\mathrm{r}} (\chi_t^{\mathrm{AR}}\varphi_t^{\mathrm{AR}} + \chi_t^\mathrm{RU}\varphi_t^\mathrm{RU})\\ 
        + s_{c} d_{\mathrm{c}} (\varrho _t^\mathrm{AR}\varphi_t^\mathrm{AR} + \varrho _t^\mathrm{RU}\varphi_t^\mathrm{RU}).
    \end{aligned}
\end{equation}
\noindent where $s_r\substack{=}\lfloor (s-1)/N_S^R \rfloor$ and $s_c\substack{=}\mod(s-1,N_S^R)$ denote row and column index of the $s$th element in the IRS, respectively.

%
\subsection{MADRL Policy for Cooperating UAV Swarm}

\par We consider each UAV as an agent and aim to adopt MADRL to facilitate cooperation among multiple UAV agents. Among various MADRL algorithms, MASAC is a state-of-the-art algorithm with a maximum entropy approach, and it also avoids massive sampling. As such, we first present the preliminaries of MASAC. Subsequently, we sequentially introduce the proposed improvements of the MASAC, which include the self-attention critic and gravity exploration.

\subsubsection{Preliminaries of MASAC}
\label{subsubsection:MASAC}

\par MASAC introduces maximum entropy and state values to enhance the exploration capabilities. Accordingly, the main idea of MASAC is to train a policy $\pi_{\theta_m}(\bm a_t^m\mid \bm s_t)$ to maximize the sum of expected rewards and the entropy objective at each visited state. The policy can be expressed as follows: 
\begin{equation}
\label{eq:best_pi}
    \pi_{\theta_m}^* = \arg \max _{\pi_{\theta_m}} \sum_{t=1}^T \mathbb{E}_\pi[r_t^m+\alpha \mathcal{H}(\pi_{\theta_m}(\cdot \mid \bm{s}_t))],
\end{equation}
\noindent where $\alpha$ is the temperature parameter determining the importance of entropy term $\mathcal{H}(\pi_{\theta_m}(\cdot \mid \bm{s}_t))$. Following this, the soft target value $y$ can be obtained as follows~\cite{haarnoja2018SAC}:
\begin{equation}
\label{eq:target}
    y= r_t^m +\gamma \mathbb{E}[Q_{\bar{\psi}_m} \left(\bm{s}_{t+1}, {\bm{a}_{t+1}}\right)-\alpha\log\pi_{\bar{\theta}_m}(\bm{a}_{t+1}^m\mid \bm{s}_{t+1})],
\end{equation}

\noindent where $Q_{\bar{\psi}_m}$ represents the evaluation function using target networks, and $\bm{a}_{t+1} = \{\bm a_{t+1}^m \sim \bar \theta _m(\cdot \mid \bm s_{t+1})\mid m \in \mathcal{M}\}$ denotes the set of actions generated by the target policy networks. To improve the critic estimation capability, we apply the temporal-difference approach to update the parameters of the value approximator by minimizing the following objective function:
\begin{equation}J_Q\left(\psi_m\right)=\sum_{m=1}^{N_M}\mathbb{E}_{(\bm{s}_{t},\bm{a}_t,r_t,\bm{s}_{t+1}) \sim \mathcal{B}} \left[\dfrac{1}{2}(Q_{\psi_m}\left({\bm{s}_t}, {\bm{a}_t}\right) - y)^2\right].
    \label{cr_update}
\end{equation}

\par Furthermore, the policy network parameters $\theta_m$ are refined through gradient-based optimization that iteratively reduces the magnitude of the following loss function, \textit{i.e.},
\begin{equation}
    \begin{aligned}
        J_\pi(\theta_m) =\mathbb{E}_{(\bm{s}_t,\bm{a}_t)\sim \mathcal{B},\bm{a}_t^m\sim \pi_{\theta_m}(\cdot\mid \bm{s}_t)} &
        [ \alpha  \log \pi_{\theta_m}(\bm{a}_t^m \mid \bm{s}_t) \\ & -Q_{{\psi}_m}(\bm{s}_t,\hat{\bm a})],
    \end{aligned}
\label{ac_update}
\end{equation} 	
\noindent where $\bm{a}_t^m$ is the action sampled from the current policy to calculate the gradient. Additionally, the action of agent $m$ in $\hat{\bm a}$ is set to $\bm{a}_t^m$, and other agent actions in $\hat{\bm a}$ are sampled from the replay buffer $\mathcal{B}$.

\par In addition, in order to stabilize the training process~\cite{liang2025rate}, the target critic and actor networks can be updated by
\begin{equation}
    \begin{aligned}
         \bar{\psi}_m=\tau \psi_m+(1-\tau) \bar{\psi}_m,\quad
         \bar{\theta}_m=\tau \theta_m+(1-\tau) \bar{\theta}_m.
    \end{aligned}
\label{tar_update}
\end{equation}
\noindent where $\tau$ denotes the soft update step size. 

\par When applying MASAC to the MDP, several critical challenges should be addressed. \textit{First}, the intricate coupling of variables creates difficulty for critic networks to accurately evaluate the state value. \textit{Second}, at the initial stages of training, experiences of collisions or moving out of boundaries restrict the ability of agents to converge to the optimum speed. \textit{Finally}, the action space of UAV agents is vast, which thus leads to insufficient exploration by MASAC. Therefore, to achieve better exploration and collaboration performance, we make the following improvements.

\begin{figure*}[t]
    \centering
    \includegraphics[width=0.95 \linewidth]{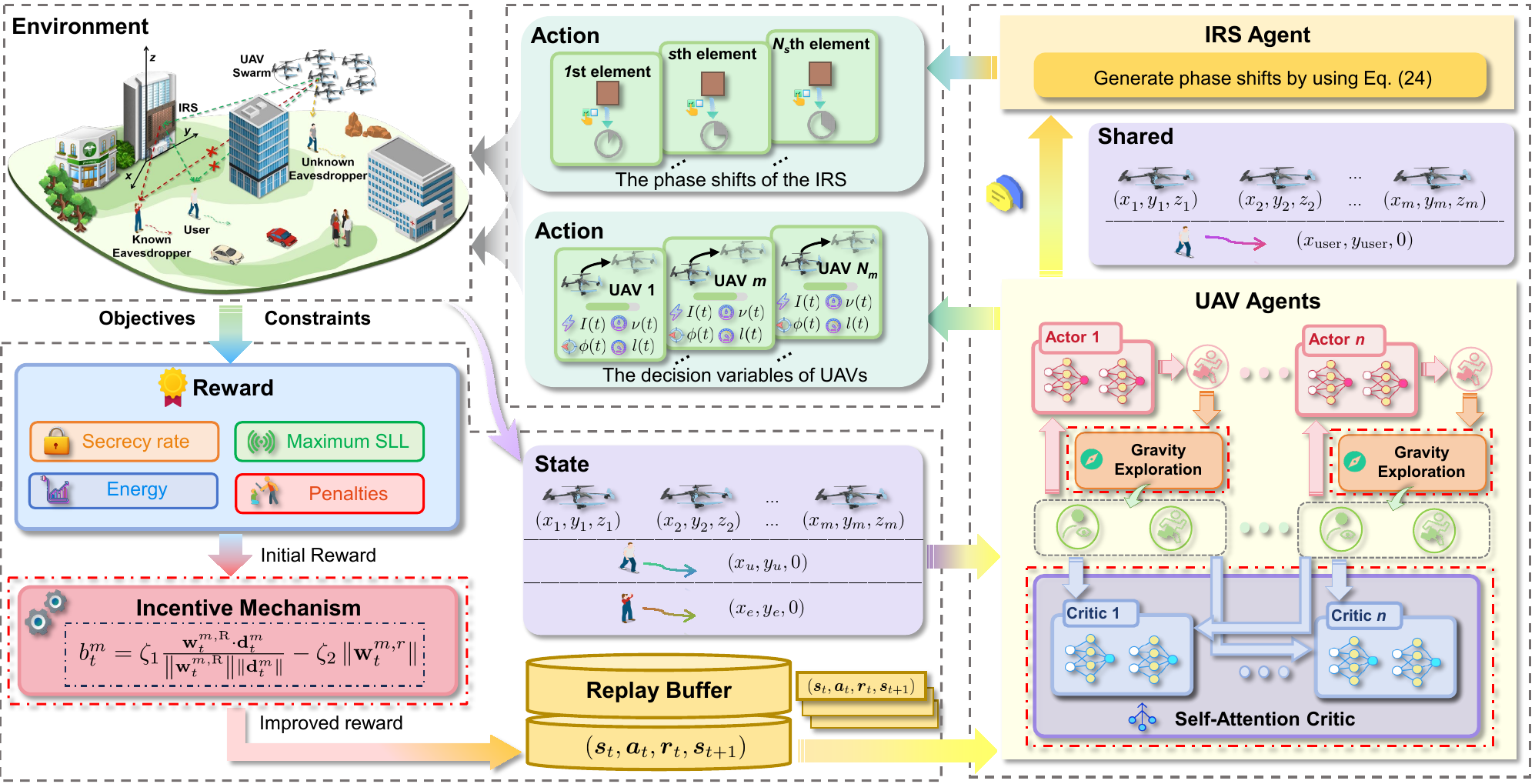}
    \caption{The proposed HMCA structure with hybrid IRS control policy and MADRL optimization method. The UAV agents based on the MADRL method and the IRS agent based on the IRS control policy.}
    \label{Fig:HMCA}
\end{figure*}

\begin{algorithm}[t]
    \KwIn{train episodes $N_{E}$, update times $N_{U}$.}
    \LinesNumbered %
    \caption{HMCA}
    \label{alg:Algorithm_HMCA} 
    \textbf{Initialization:} Initialize the actor $\theta_m$, target actor $\bar{\theta}_m =\theta_m$, critic $\psi_m$, target critic $\bar{\psi}_m = \psi_m$ and replay buffer $\mathcal{B}$ for each agent;\\
    \For{$ n=1 $ to $N_E$}
    {
        Initialize and observes the state $\bm{s}_t$;\\
        \For{$ t=1 $ to $ T $}
        {
            Each UAV sample obtained $\bm{a}_t^m$ by sampling distribution $ \pi_{\theta_m}(\cdot\mid \bm{s}_t) $;\\
            Add gravity noise to $\bm{a}_t^m$ using Eq.~\eqref{eq:gravity_exploration} and obtain the noised action $\hat{\bm a}_t^m$;\\
            Generate the phase shifts of IRS by Eq.~\eqref{eq:irs_op_method};\\
            UAVs and the IRS perform actions and receive the reward $r_t^m$;\\
            Obtain the next state $\bm{s}_{t+1}$;\\
            Set equivalent rewards $\hat{\bm r}_t =\bm{r}_t + \bm{b}_t$;\\
            Union and store the experience tuple $(\bm{s}_t, \hat{\bm{a}}_t,\hat{\bm r}_t,\bm{s}_{t+1})$ in $\mathcal{B}$;\\
            Update state $\bm s_{t} \leftarrow \bm s_{t+1}$;\\
            
            \For{$i=1$ to $N_{U}$}
            {
                Randomly sample a mini-batch of experiences $ (\bm{s},\bm{a},\bm{r},\bm{s}^\prime) $ from $ \mathcal{B} $;\\        
                \For{$m=1$ to $N_{M}$}{
                    Calculate $Q_{\psi_m}(\bm s,\bm a)$ and update $\psi_m$ using Eqs.~\eqref{cr_update} and~\eqref{eq:attention_Q};\\
                    Update $\theta_m$ using Eq.~\eqref{ac_update};\\
                    Update $\bar \psi_m$ and $\bar \theta_m$ using Eq.~\eqref{tar_update};\\
                }
            }
        }
    }		
    \KwOut{Policies $\{\pi_{\theta_1},\ldots,\pi_{\theta_m},\ldots,\pi_{\theta_{N_M}}\}$;\\}
\end{algorithm}

%
\subsubsection{Self-Attention Critic}

\par In the MDP, different UAV agents seek to cooperate and have non-conflicting interests. Under this circumstance, the multilayer perceptron (MLP) overlooks the interaction information among the UAV agents~\cite{vaswani2017attention}. Thus, we use a critic network based on a self-attention mechanism to replace the MLP, thereby improving collaborative reward evaluation. As a result, the Q-function $Q_{\psi_m}(\bm{s}, \bm{a})$ combines the weighted actions of other UAV agents, which is given as follows:
\begin{equation}
    Q_{\psi_m}(\bm{s}, \bm{a}) = f_m(g_m(\bm{s}, \bm{a}_t^m), \bm{x}_m),
    \label{eq:attention_Q}
\end{equation}
\noindent where $f_m$ is a two-layer MLP and $g_m$ is a one-layer MLP, while $\bm{x}_m$ is the attention interaction information of the agent $m$ with other agents. To obtain $\bm{x}_m$, we use a set of parameters $\bm{W}_q$, $\bm{W}_k$ and $\bm{W}_v$ to accomplish the relevant transformations. According to~\cite{vaswani2017attention}, the query $q_m$, key $k_m$ and value $v_m$ can be expressed as follows:
\begin{equation}
    q_m = g_m(\bm{s},\bm{a}_t^m)\bm{W}_q, \ k_m= \bm{a}_t^m  \bm{W}_k, \ v_m=  \bm{a}_t^m\bm{W}_v.
    \label{eq:attention_query}
\end{equation}

\par Subsequently, the attention score of the agent $m$ to the agent $n$ is given by $s_{m,n} = \mathrm{softmax}(q_m k_n^T/\sqrt{d_{k}})$, where $ d_{k} $ is the dimension of the $k_m$~\cite{vaswani2017attention}. Consequently, $ \bm{x}_m $ can be obtained by $\bm x_m = \sum_{m \ne n } s_{m,n} v_n$.

\par Note that the critic network is not involved during execution, which means that our method still maintains the advantages of decentralization. Furthermore, the weight matrices $\bm{W}_q$, $\bm{W}_k$, and $\bm{W}_v$ are shared across all agents, thus enhancing the learning efficiency~\cite{iqbal2019actorMAAS}.

%
\subsubsection{Gravity Exploration Scheme}

\par During the training process, UAV agents are prone to out-of-bounds events or collisions in the early stages, thus leading to slow convergence. According to~\cite{Ma2024}, the energy-optimal speed for a UAV can be calculated via UAV parameters and defined as $v_\mathrm{me}$. Therefore, we consider applying a gravity exploration scheme to the horizontal speed of UAVs during training.

\par Specifically, for the horizontal speed $v_m$ obtained from $ \pi_{\theta_m} $, we add a gravity noise to make $v_m$ closer to the ideal speed $ v_\mathrm{me} $. Furthermore, a self-adaptive strategy is employed to ensure the convergence of the model in the later stages of training. As such, the noised horizontal speed $ \tilde{v}_m$ can be obtained as follows:
\begin{equation}
\label{eq:gravity_exploration}
    \tilde{v}_m = \mathrm{clip}\left(\zeta \times v_m + (1-\zeta)\times v_0,v_{min},v_{max}\right),
\end{equation}

\noindent where $\zeta = {n}/N_{E}$ represents the current training progress, and $ v_0 \sim \mathcal{M}( v_\mathrm{me}, \sigma_0^2)$ denotes the variable drawn from a Gaussian distribution with mean $ v_\mathrm{me} $ and variance $\sigma_0^2$. Additionally, the $ \mathrm{clip} $ operation is used to ensure the output horizontal speed $\tilde{v}_m$ is in the legitimate range $ [v_{min},v_{max}]$.

%

\subsection{HMCA Framework and Steps}

\par According to the aforementioned methods, we summarize an HMCA to solve the formulated problem shown in Eq.~\eqref{problem}. The main steps of the HMCA are shown in Algorithm~\ref{alg:Algorithm_HMCA}, and the structure is presented as follows. 

\par As illustrated in Fig.~\ref{Fig:HMCA}, HMCA consists of an IRS agent and UAV agents. Specifically, the IRS agent makes actions by using the policy proposed in Section~\ref{subsection:irs_control_policy}, while UAV agents are controlled by the proposed enhanced MASAC-based method. Each UAV agent is with four neural networks with parameters $\theta_m$, $\psi_m$, $\bar \theta_m$, and $\bar \psi_m$.

\par During the training, both the IRS and UAV agents interact with the environment and store the experience tuple in the replay buffer $\mathcal{B}$. Specifically, the UAV agent first obtains a raw action $a_t^m$ using the policy network. By applying Eq.~\eqref{eq:gravity_exploration}, the UAV agent obtains a new action with gravity noise $\hat{\bm a}_t^m$. Subsequently, the IRS agent receives the information from the UAV, and performs action via Eq.~\eqref{eq:irs_op_method}. Following this, the environment will return a reward to both IRS and UAV agents. 

\par During the execution stage, the HMCA can be deployed in a decentralized manner since the agents are independent. Additionally, in practice, IRS can use a smart controller to obtain the actions of UAVs, thereby turning their phase shifts in real time~\cite{IRS_wuqingqing2}.

\subsection{Complexity Analyses of HMCA}

\par The computational and space complexity of HMCA during the training and execution phases are analyzed in the following sections. 

\par The computational complexity of HMCA during the training phase is $\mathcal{O}(N_ETN_UN_M(N_Md(|\boldsymbol{s}| + |\boldsymbol{a}|) + {N_M}^2d + N_Md(2d) + |\boldsymbol{\theta_m}| + |\boldsymbol{\bar \theta_m}| + |\boldsymbol{\psi_m}| + |\boldsymbol{\bar \psi_m}|))$, which is given by

\begin{itemize}

\item \textit{Network Initialization:} During this initial setup, we need to configure parameter values for all $N_M$ neural networks of UAV agents. The computational resources required for this initialization can be quantified as $\mathcal{O}(N_M(|\boldsymbol{\theta_m}| + |\boldsymbol{\bar \theta_m}| + |\boldsymbol{\psi_m}| + |\boldsymbol{\bar \psi_m}|))$, with $| \cdot | $ indicating the parameter count within each respective network structure.

\item \textit{Action Transition:} For executing the policy decisions, the algorithm needs to process actor network outputs and convert them into appropriate actions, resulting in computational demands of $\mathcal{O}(N_ETN_M)$. In this expression, $N_E$ represents the total episode count during training, while $T$ corresponds to the timestep quantity per training episode.

\item \textit{Reward Calculation and State Transitions:} The computational complexity of reward calculation and state transitions is $\mathcal{O}(N_ETN_MV)$, where $V$ represents the complexity of interacting with the environment.

\item \textit{Network Update:} The update phase consists of two main parts. First, it involves calculating the Q-value function with the self-attention mechanism and subsequently updating the parameters of the critic network accordingly. Following this, it proceeds to update the parameters of the actor network and then performs a soft update of the parameters of the target network. 
The computational complexity of the critic network with a self-attention mechanism includes three main components. State and action encoding has a complexity of $\mathcal{O}(N_Md(|\boldsymbol{s}| + |\boldsymbol{a}|))$, where $|\boldsymbol{s}|$ and $|\boldsymbol{a}|$ denote the dimensions of the state and action spaces, and $d$ is the hidden layer dimension. The self-attention mechanism has a complexity of $\mathcal{O}({N_M}^2d)$. Furthermore, Q value calculation has a complexity of $\mathcal{O}(N_Md(2d))$. Overall, the complexity of the first stage is $\mathcal{O}(N_ETN_UN_M(N_Md(|\boldsymbol{s}| + |\boldsymbol{a}|) + {N_M}^2d + N_Md(2d) + |\boldsymbol{\psi_m}|))$, where $N_U$ denotes the number of updates per agent per training round. Therefore, the complexity of this phase is calculated as $\mathcal{O}(N_ETN_UN_M(N_Md(|\boldsymbol{s}| + |\boldsymbol{a}|) + {N_M}^2d + N_Md(2d) + |\boldsymbol{\theta_m}| + |\boldsymbol{\bar \theta_m}| + |\boldsymbol{\psi_m}| + |\boldsymbol{\bar \psi_m}|))$. 

\end{itemize}

\par Furthermore, the space complexity during HMCA training can be expressed as $\mathcal{O}(N_M(|\boldsymbol{\theta_m}| + |\boldsymbol{\bar \theta_m}| + |\boldsymbol{\psi_m}| + |\boldsymbol{\bar \psi_m}|) + |\boldsymbol{\mathcal{B}}|(2|\boldsymbol{s}| + |\boldsymbol{a}| + 1)$, with $|\boldsymbol{\mathcal{B}}|$ indicating the replay buffer capacity. This reflects the memory needed to store all network weights across agents and the experience collection mechanism that maintains state-action-reward-nextstate sequences for learning.

\par During the execution phase, the computational complexity of HMCA is $\mathcal{O}(N_E N_M)$, which stems from action selection and transition according to the current state using the actor network. Moreover, the space complexity during the execution phase is $\mathcal{O}(N_M|\boldsymbol{\theta_m}|)$ since the parameters of the actor network need to be stored for action selection, which thus enables efficient execution.

%
\section{Simulation Results}\label{sec:simulation}

\par In this section, we verify the performance of the HMCA and compare it with the baselines. We begin by demonstrating the convergence of the proposed HMCA. Subsequently, we conduct a comparative analysis with baselines. Finally, we present some visualization results.

\subsection{Simulation Setups}

\par We set the number of UAVs with in UAV swarm $N_M$ as 8, and these UAVs are restricted in a flyable area of 100 m $\times$ 100 m~\cite{LiCb3}. Additionally, the minimum and maximum altitudes of UAVs are 75 m and 95 m, respectively, while the minimum distance between any two UAVs is set to 0.5 m~\cite{LiCb3}. Furthermore, the height of IRS is set to 20 m, and the number of the passive reflecting elements $N_S$ is set to 60. Moreover, the carrier frequency, transmit power $P_m$, path loss exponent $\alpha_r$, path loss index $\alpha_d$, bandwidth, and total noise power spectral density are set as 2.4 GHz, 0.1 W, 2.7, 3.6, 20 MHz, and -155 dBm/Hz~\cite{LiCb3}, respectively. Finally, the UAV propulsion energy parameters $m_M$, $v_{tip}$, $ v_0$, $\rho$, $A$, $d_0$, and $s$ are set to 2 kg, 120 m/s, 4.03 m/s, 1.225 kg/$\mathrm m^3$, 0.503 $\mathrm{m}^3$, 0.6 and 0.05~\cite{Ma2024}, respectively. 

\par To obtain a more accurate algorithm training convergence curve, the average episode reward is used as the performance metric. After comprehensive parameter tuning, the number of nodes in the hidden layer, target network update step $\tau$, batch size $N_{B}$, learning rate, number of episodes $N_E$, temperature parameter $\alpha$, and discount factor are set to 256, 0.005, 256, 0.0003, 2000, 0.01 and 0.95, respectively. The ReLU function~\cite{Mao2025} is employed as the activation function, whereas the Tanh function~\cite{Mao2025} is utilized to scale the output, which thus ensures appropriate action values. 


\par To highlight the superiority of the MADRL approach in HMCA, we introduce the following baselines, and these methods adopt the same parameters as mentioned above. Note that all these methods integrate the previously discussed IRS control policy for training purposes, which thus ensures consistent comparison conditions. The subsequent section will present a detailed demonstration of the performance exhibited by this IRS control policy, thereby providing comprehensive evidence of its effectiveness.

\begin{figure}[t]
	\centering
	\subfloat[]{\includegraphics[width=0.99\linewidth]{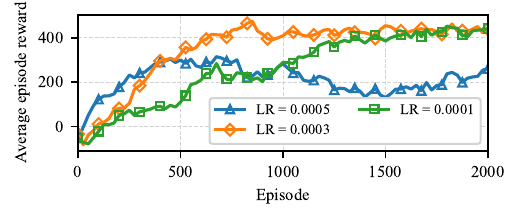}}
    \label{fig:convergence_lr}\hfill 
    \\
    \centering
    \subfloat[]{\includegraphics[width=0.99\linewidth]{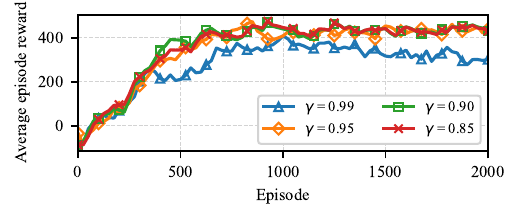}}
    \label{fig:convergence_gamma}\hfill
    \caption{Hyperparameter tuning results. (a) Different learning rates. (b) Different discount factors. }
    \label{Fig:convergences_lr_gamma}
\end{figure}

\begin{itemize}
\item \textit{Multi-Agent Proximal Policy Optimization (MAPPO):} MAPPO is an on-policy method~\cite{Cheng2025} that utilizes a trust region-based approach to ensure stable policy improvements. Moreover, MAPPO employs a clipped surrogate objective function which effectively prevents destructively large policy updates, thus enhancing training stability in multi-agent environments.

\item \textit{Multi-Agent Deep Deterministic Policy Gradient (MADDPG):} MADDPG is a classical MADRL approach based on the architecture of centralised training and decentralised execution (CTDE)~\cite{pmlr-v80-haarnoja18bMADDPG}. MADDPG extends the actor-critic framework to multi-agent settings and excels in handling continuous action spaces. MADDPG also incorporates a centralized critic that can access all observations and actions of agents during training, thereby facilitating more informed policy updates.

\item \textit{MASAC:} MASAC utilizes the CTDE architecture similar to MADDPG. The details of this approach are elaborated in Section~\ref{subsubsection:MASAC}. The implementation adheres to the principles of entropy-based exploration while maintaining centralized training, which enables it to effectively balance exploration and exploitation in complex multi-agent scenarios.

\item \textit{Single-Agent Learning (SAL)}: SAL is based on soft actor-critic (SAC) algorithm, and used to demonstrate the effectiveness of the multi-agent approach. In this method, each agent adopts a SAC algorithm independently without coordination mechanisms. Different from MASAC, SAL does not incorporate the CTDE architecture~\cite{SAC2}, which results in less coordinated behavior among the agents and limited ability to model other agent behaviors. 

\item \textit{Random}: For this baseline approach, each agent takes a random action during each timeslot without any learning or adaptation mechanism. This method serves as a lower bound for performance evaluation, thus providing context for the improvements achieved by the learning-based methods. The random policy demonstrates the performance achievable without any intelligent decision-making, which highlights the value of structured learning approaches in complex multi-agent tasks.

\end{itemize}

%

\subsection{Hyperparameter Tuning}

\par First, we investigate the impact of the learning rate on the performance of the algorithm. As illustrated in Fig.~\ref{Fig:convergences_lr_gamma}(a), HMCA demonstrates accelerated convergence when the learning rate increases. However, excessively rapid convergence may direct the policy towards local optimization rather than global solutions. Therefore, it is crucial to establish an appropriate balance when setting the learning rate. Thus, we select a fixed learning rate of 0.0003 for subsequent simulations, which provides a suitable balance with moderate convergence speed and stable learning characteristics.

\par Moreover, in Fig.~\ref{Fig:convergences_lr_gamma}(b), we assess the influence of the discount factor on the algorithm performance. As can be seen, an exceedingly high discount factor impedes network convergence. To ensure that the evaluation network considers long-term gains while simultaneously maintaining learning stability, we utilize 0.95 as the discount factor for subsequent simulations, thereby achieving an appropriate balance between current and long-term rewards.

\subsection{Optimization Results}

\begin{figure*}[t]
\begin{minipage}[t]{1\linewidth}
  \centering
        \subfloat[]{\includegraphics[width=.295\linewidth]{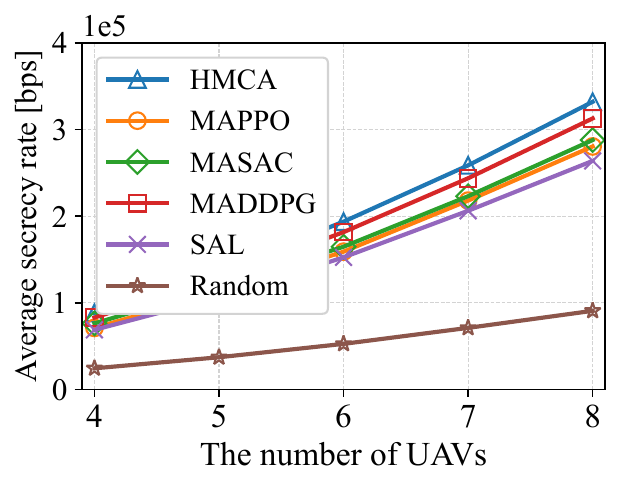}}\hfill
	\subfloat[]{\includegraphics[width=.32\linewidth]{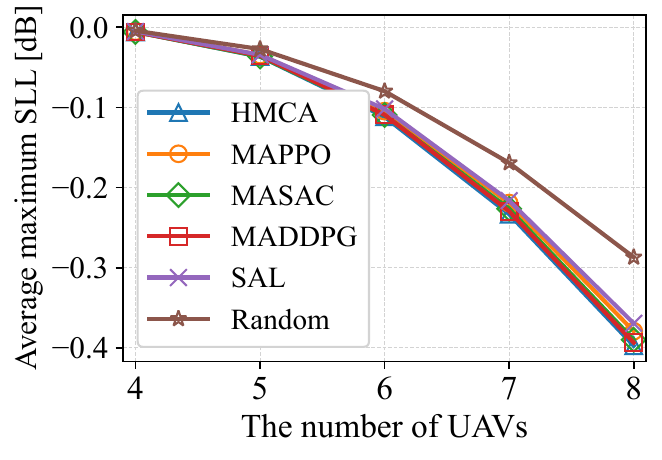}}\hfill
	\subfloat[]{\includegraphics[width=.305\linewidth]{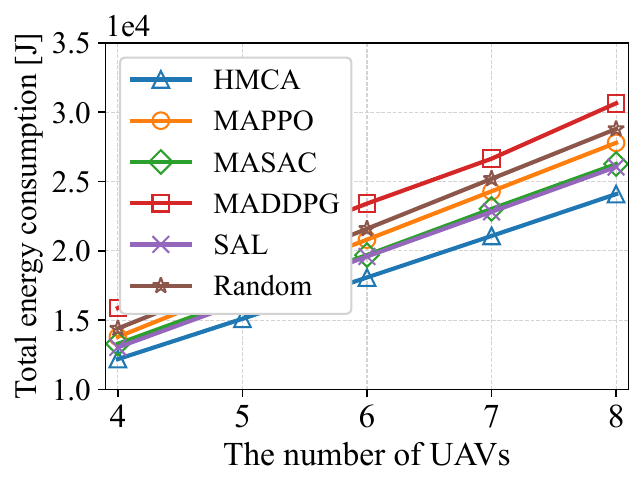}}\hfill
    \caption{Optimization results obtained by different methods under varying numbers of UAVs. (a) Average secrecy rate. (b) Average total maximum SLL. (c) Total energy consumption obtained by different methods.}
    \label{Fig:convergences}
\end{minipage}
\end{figure*}

%
\subsubsection{Optimization Objective Results}

\par Figs.~\ref{Fig:convergences}(a),~\ref{Fig:convergences}(b), and~\ref{Fig:convergences}(c) present comparative performance analyses across different algorithms for three optimization objectives, \textit{i.e.}, average secrecy rate ($f_1$), average total maximum SLL ($f_2$), and total UAV energy consumption ($f_3$), respectively. As can be seen, the results demonstrate that HMCA consistently outperforms all baseline approaches, with particularly significant advantages in secrecy rate optimization. Furthermore, all DRL-based methods achieve substantially better results than the random approach, which confirms the effectiveness and suitability of DRL frameworks for addressing our complex optimization problem. 

\par Moreover, the figures also reveal important scaling properties of the system. Specifically, as the number of UAVs increases, both the average secrecy rate and the maximum SLL show marked improvement, indicating enhanced communication security. Notably, this security enhancement comes with only a linear increase in energy consumption relative to UAV count. These observations support an important system design insight, \textit{i.e.}, deploying additional UAVs represents an efficient strategy for significantly improving secure communication performance while incurring proportionally minimal energy overhead.

%
\subsubsection{Convergence Results}

\begin{figure}[b]
\centering
\begin{minipage}[h]{1\linewidth}
  \centering
  \includegraphics[width=3.5 in]{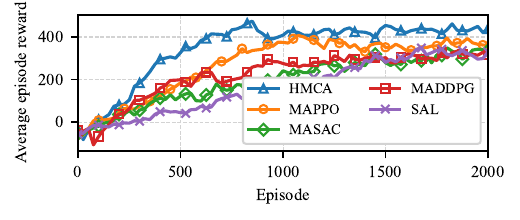}
  \caption{Convergence curve obtained by different methods.}
  \label{fig:convergence}
\end{minipage}
\end{figure}

\par Fig.~\ref{fig:convergence} shows the learning convergence curve obtained by different methods. As can be seen, the proposed HMCA achieves the highest average episode reward and outperforms other methods in convergence speed, requiring approximately 20\% fewer training episodes to reach stable performance compared to baseline methods. This superior performance can be attributed to the gravity exploration scheme, which enhance the exploration capabilities of the UAV agents, thereby achieving fast convergence. This accelerated convergence represents the advantage in practical applications, particularly in dynamic environments where rapid adaptation is essential. In such cases, this characteristic facilitates quicker fine-tuning when system parameters change, thus enhancing the overall adaptability and responsiveness of the UAV network in uncertain or evolving operational contexts.

\par Moreover, the SAL approach, while achieving sub-optimal results in energy consumption, exhibits significantly poorer performance regarding secrecy rate and maximum SLL. This performance disparity highlights the inherent advantage of the joint state-action evaluation methodology employed in MADRL approaches, which substantially enhances collaborative decision-making capabilities. By contrast, the SAL approach tends to prioritize optimization of individual agent energy metrics while struggling to effectively evaluate collaborative rewards. In addition, traditional CTDE-based multi-agent approaches encounter notable convergence challenges when operating within the high-dimensional joint state-action spaces characteristic of multi-UAV environments. These dimensionality issues inherently constrain the exploration capabilities of agent, resulting in diminished convergence performance. Thus, the superior results achieved by HMCA in this context clearly demonstrate the efficacy of its attention mechanism, which successfully addresses these limitations by enabling more efficient state representation and inter-agent coordination while navigating complex collaborative tasks.

\subsubsection{Policy Evaluation and Visualization Results}

\par
\begin{figure}
\centering
\begin{minipage}[h]{1\linewidth}
  \centering
  \includegraphics[width=3.3 in]{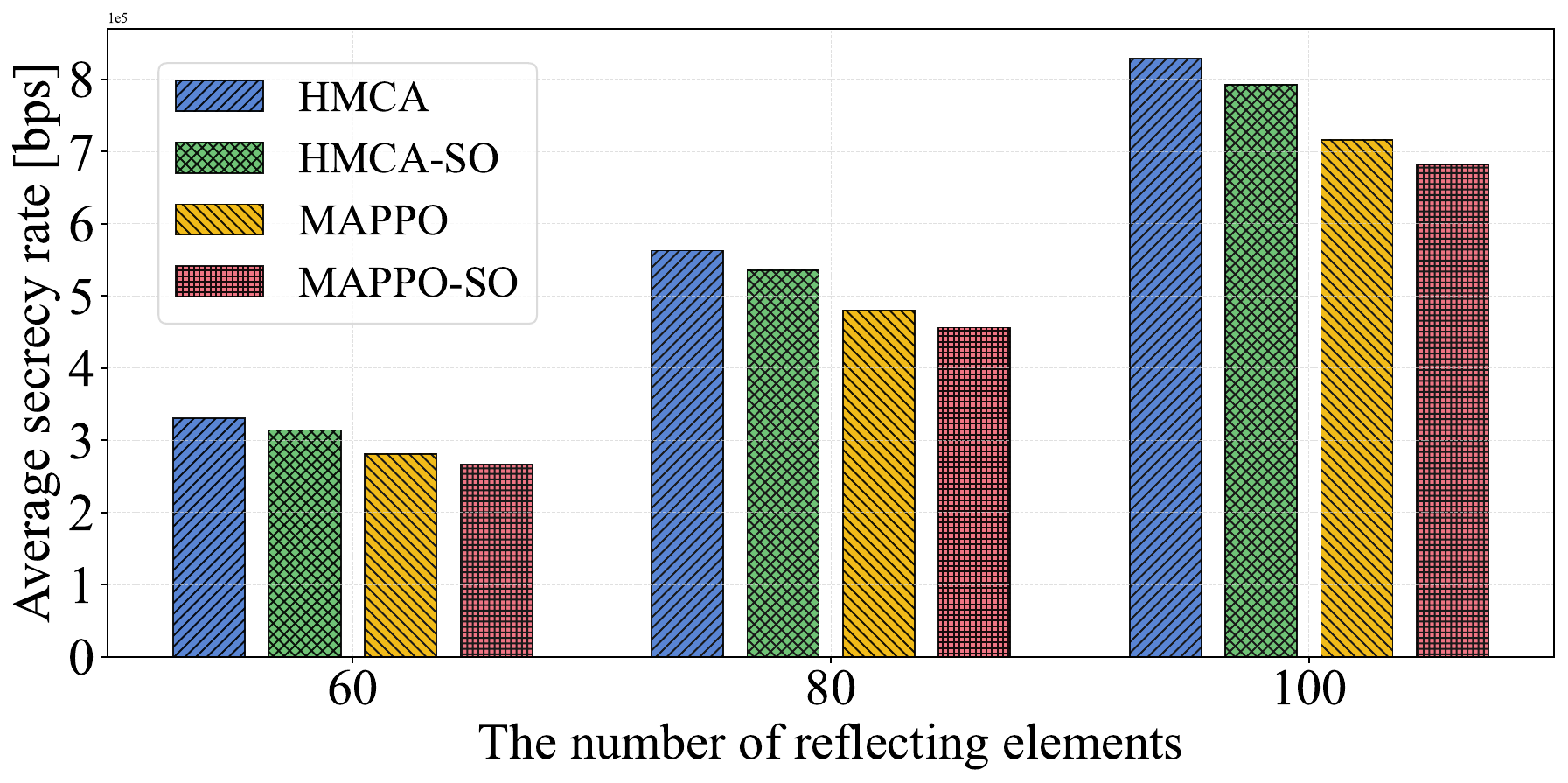}
  \caption{IRS policy comparisons under different numbers of elements.}
  \label{fig:irs_irsnum}
\end{minipage}
\end{figure}

 To evaluate the effectiveness of the employed IRS controlling policy, we conduct a comparative analysis against the method proposed in~\cite{wu2019beamforming}. Fig.~\ref{fig:irs_irsnum} illustrates the secrecy rate performance of both IRS control policies when integrated with HMCA and MAPPO algorithms, across varying numbers of UAVs and IRS reflecting units. The results clearly demonstrate the superior performance of our UAV and user location-based control policy, which consistently achieves higher secrecy rates across all configurations. Notably, this performance advantage becomes increasingly pronounced as the number of reflecting units increases, highlighting the scalability and effectiveness of our approach in complex communication environments.

\par
\begin{figure}
\centering
\begin{minipage}[h]{0.9\linewidth}
    \centering
  \includegraphics[width=3.35in]{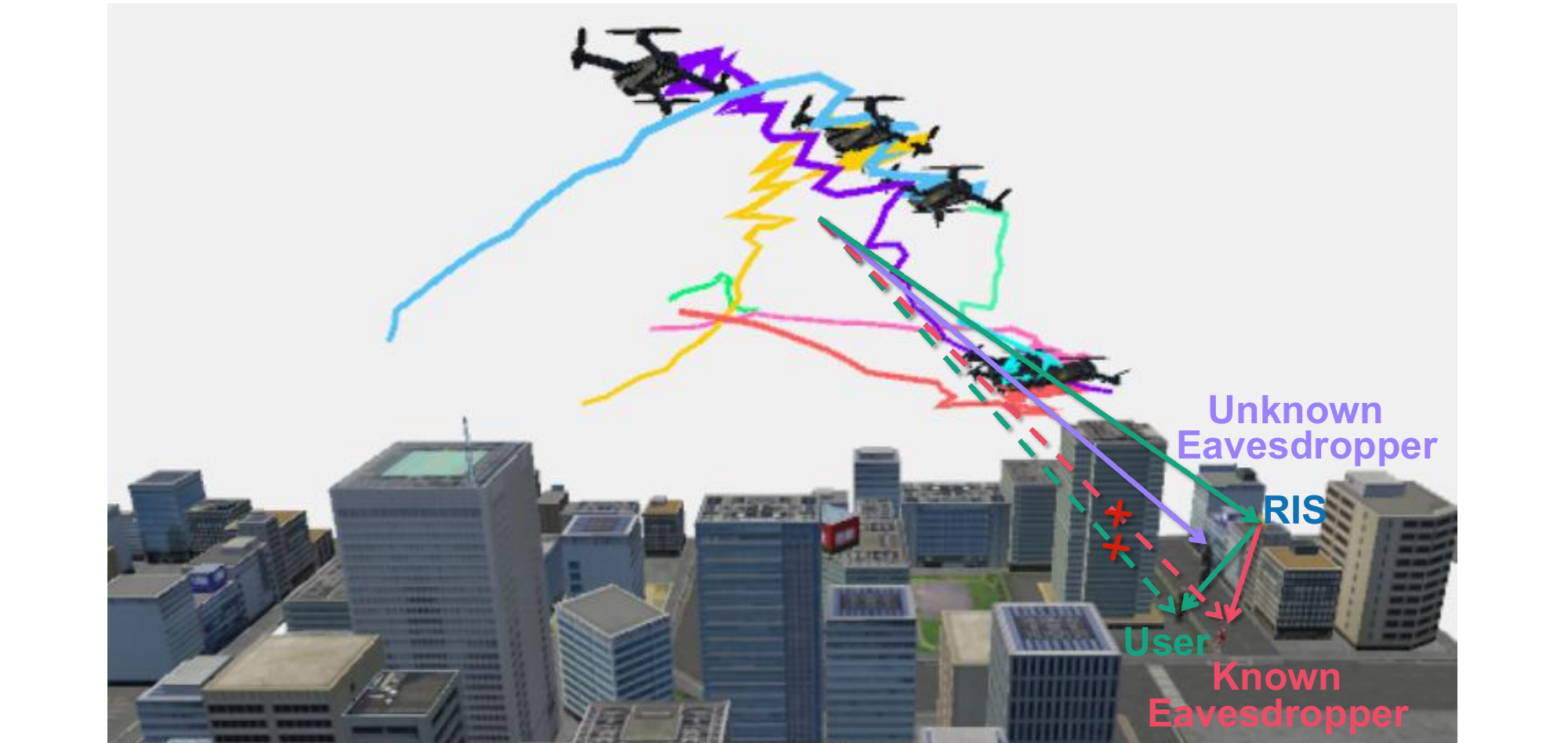}
  \caption{The trajectories of UAVs obtained by HMCA.}
  \label{fig:tra}
\end{minipage}

\end{figure}

 Fig.~\ref{fig:tra} provides a visualization of UAV movement patterns, which displays the trajectories of UAV swarm. The visualization reveals that UAVs systematically migrate toward the IRS location over time, which confirms the effectiveness of our proposed incentive mechanism. This behavior demonstrates how the mechanism successfully guides UAV exploration patterns toward optimal positions that leverage IRS capabilities, thereby enhancing the secure performance of communication.

\section{Conclusion}\label{sec:conclusion}

\par This paper has studied the PLS in UAV swarm-IRS collaborative secure communication system. We have formulated a JCPBOP to increase secrecy rate, reduce SLL, and minimize energy consumption. To address this problem, we have proposed a novel HMCA framework. Simulation results have demonstrated that HMCA significantly outperforms baseline approaches, including MAPPO, MADDPG, MASAC, and SAL across all optimization objectives. Moreover, the advantage of HMCA becomes more pronounced as the number of UAVs increases, with the average secrecy rate and maximum SLL metrics showing marked improvement while energy consumption increases only linearly with UAV count. For future work, an important direction involves extending the current IRS framework to simultaneously transmit and reflect reconfigurable intelligent surfaces, thereby potentially enhancing system flexibility and coverage through the dual capabilities of signal transmission and reflection.




\bibliographystyle{IEEEtran}
\bibliography{reference}

\begin{thebibliography}{10}
\providecommand{\url}[1]{#1}
\csname url@samestyle\endcsname
\providecommand{\newblock}{\relax}
\providecommand{\bibinfo}[2]{#2}
\providecommand{\BIBentrySTDinterwordspacing}{\spaceskip=0pt\relax}
\providecommand{\BIBentryALTinterwordstretchfactor}{4}
\providecommand{\BIBentryALTinterwordspacing}{\spaceskip=\fontdimen2\font plus
\BIBentryALTinterwordstretchfactor\fontdimen3\font minus \fontdimen4\font\relax}
\providecommand{\BIBforeignlanguage}[2]{{%
\expandafter\ifx\csname l@#1\endcsname\relax
\typeout{** WARNING: IEEEtran.bst: No hyphenation pattern has been}%
\typeout{** loaded for the language `#1'. Using the pattern for}%
\typeout{** the default language instead.}%
\else
\language=\csname l@#1\endcsname
\fi
#2}}
\providecommand{\BIBdecl}{\relax}
\BIBdecl

\bibitem{lai2024irs}
L.~Wei, H.~Kang, G.~Sun, J.~Li, J.~Wang, and D.~Niyato, ``Irs-assisted uav secure communications via joint collaborative and passive beamforming,'' in \emph{Proc. IEEE MSN}, 2024.

\bibitem{yuan2025ground}
W.~Yuan, Y.~Cui, J.~Wang, F.~Liu, G.~Sun, T.~Xiang, J.~Xu, S.~Jin, D.~Niyato, S.~Coleri \emph{et~al.}, ``From ground to sky: Architectures, applications, and challenges shaping low-altitude wireless networks,'' \emph{arXiv preprint arXiv:2506.12308}, 2025.

\bibitem{SunTMC}
G.~Sun, L.~He, Z.~Sun, Q.~Wu, S.~Liang, J.~Li, D.~Niyato, and V.~C.~M. Leung, ``Joint task offloading and resource allocation in aerial-terrestrial {UAV} networks with edge and fog computing for post-disaster rescue,'' \emph{{IEEE} Trans. Mob. Comput.}, vol.~23, no.~9, pp. 8582--8600, 2024.

\bibitem{Long2024}
Y.~Long, S.~Zhao, S.~Gong, B.~Gu, D.~Niyato, and X.~Shen, ``Aoi-aware sensing scheduling and trajectory optimization for multi-uav-assisted wireless backscatter networks,'' \emph{{IEEE} Trans. Veh. Technol.}, vol.~73, no.~10, pp. 15\,440--15\,455, 2024.

\bibitem{Hoang2025TON}
L.~T. Hoang, C.~T. Nguyen, H.~D. Le, and A.~T. Pham, ``Adaptive {3D} placement of multiple {UAV}-mounted base stations in {6G} airborne small cells with deep reinforcement learning,'' \emph{{IEEE} Trans. Netw.}, 2025, early access, doi: 10.1109/TON.2025.3552097.

\bibitem{jin2025predictive}
H.~Jin, J.~Wu, W.~Yuan, R.~Ruan, J.~Wang, D.~Niyato, D.~I. Kim, and A.~Jamalipour, ``Predictive control over lawn: Joint trajectory design and resource allocation,'' \emph{arXiv preprint arXiv:2507.02374}, 2025.

\bibitem{Sun2025}
G.~Sun, J.~Xiao, J.~Li, J.~Wang, J.~Kang, D.~Niyato, and S.~Mao, ``Aerial reliable collaborative communications for terrestrial mobile users via evolutionary multi-objective deep reinforcement learning,'' \emph{{IEEE} Trans. Mob. Comput.}, vol.~24, no.~7, pp. 5731--5748, 2025.

\bibitem{Sun2022}
G.~Sun, J.~Li, A.~Wang, Q.~Wu, Z.~Sun, and Y.~Liu, ``Secure and energy-efficient {UAV} relay communications exploiting collaborative beamforming,'' \emph{{IEEE} Trans. Commun.}, vol.~70, no.~8, pp. 5401--5416, 2022.

\bibitem{Zheng2024}
X.~Zheng, G.~Sun, J.~Li, S.~Liang, Q.~Wu, M.~Yin, D.~Niyato, and V.~C.~M. Leung, ``Reliable and energy-efficient communications via collaborative beamforming for {UAV} networks,'' \emph{{IEEE} Trans. Wirel. Commun.}, vol.~23, no.~10, pp. 13\,235--13\,251, 2024.

\bibitem{ZhangCb2}
C.~Zhang, G.~Sun, Q.~Wu, J.~Li, S.~Liang, D.~Niyato, and V.~C.~M. Leung, ``{UAV} swarm-enabled collaborative secure relay communications with time-domain colluding eavesdropper,'' \emph{{IEEE} Trans. Mob. Comput.}, vol.~23, no.~9, pp. 8601--8619, 2024.

\bibitem{LiCb3}
J.~Li, G.~Sun, L.~Duan, and Q.~Wu, ``Multi-objective optimization for {UAV} swarm-assisted {IoT} with virtual antenna arrays,'' \emph{{IEEE} Trans. Mob. Comput.}, vol.~23, no.~5, pp. 4890--4907, 2024.

\bibitem{IRS_wuqingqing}
Q.~Wu and R.~Zhang, ``Towards smart and reconfigurable environment: Intelligent reflecting surface aided wireless network,'' \emph{{IEEE} Commun. Mag.}, vol.~58, no.~1, pp. 106--112, 2020.

\bibitem{IRS_wuqingqing2}
Q.~Wu, S.~Zhang, B.~Zheng, C.~You, and R.~Zhang, ``Intelligent reflecting surface-aided wireless communications: {A} tutorial,'' \emph{{IEEE} Trans. Commun.}, vol.~69, no.~5, pp. 3313--3351, 2021.

\bibitem{Saif2024}
M.~Saif and S.~Valaee, ``Improving connectivity of {RIS}-assisted {UAV} networks using {RIS} partitioning and deployment,'' in \emph{Proc. {IEEE} {VTC}}, 2024, pp. 1--6.

\bibitem{Li2024a}
B.~Li, J.~Liao, W.~Wu, and Y.~Li, ``Intelligent reflecting surface assisted secure computation of wireless powered {MEC} system,'' \emph{{IEEE} Trans. Mob. Comput.}, vol.~23, no.~4, pp. 3048--3059, 2024.

\bibitem{Yi2025}
D.~Yi and H.~Zhang, ``Performance analysis of {IRS}-assisted networks with near- and far-field effects,'' \emph{{IEEE} Trans. Veh. Technol.}, no.~4, pp. 6739--6744, Apr. 2025.

\bibitem{Ning}
Z.~Ning, Y.~Yang, X.~Wang, Q.~Song, L.~Guo, and A.~Jamalipour, ``Multi-agent deep reinforcement learning based {UAV} trajectory optimization for differentiated services,'' \emph{{IEEE} Trans. Mob. Comput.}, vol.~23, no.~5, pp. 5818--5834, 2024.

\bibitem{Hao}
H.~Hao, C.~Xu, W.~Zhang, S.~Yang, and G.~Muntean, ``Joint task offloading, resource allocation, and trajectory design for multi-{UAV} cooperative edge computing with task priority,'' \emph{{IEEE} Trans. Mob. Comput.}, vol.~23, no.~9, pp. 8649--8663, 2024.

\bibitem{zhao2022multidrl1}
N.~Zhao, Z.~Ye, Y.~Pei, Y.~Liang, and D.~Niyato, ``Multi-agent deep reinforcement learning for task offloading in {UAV}-assisted mobile edge computing,'' \emph{{IEEE} Trans. Wirel. Commun.}, vol.~21, no.~9, pp. 6949--6960, 2022.

\bibitem{Ye}
Z.~Ye, K.~Wang, Y.~Chen, X.~Jiang, and G.~Song, ``Multi-{UAV} navigation for partially observable communication coverage by graph reinforcement learning,'' \emph{{IEEE} Trans. Mob. Comput.}, vol.~22, no.~7, pp. 4056--4069, 2023.

\bibitem{Tariq2024}
Z.~U.~A. Tariq, E.~Baccour, A.~Erbad, and M.~Hamdi, ``Reinforcement learning for resilient aerial-{IRS} assisted wireless communications networks in the presence of multiple jammers,'' \emph{{IEEE} Open J. Commun. Soc.}, vol.~5, pp. 15--37, 2024.

\bibitem{yang2022onlineGMRMM}
Z.~Yang, S.~Bi, and Y.~A. Zhang, ``Online trajectory and resource optimization for stochastic {UAV}-enabled {MEC} systems,'' \emph{{IEEE} Trans. Wirel. Commun.}, vol.~21, no.~7, pp. 5629--5643, 2022.

\bibitem{Li2024tmc}
J.~Li, G.~Sun, L.~Duan, and Q.~Wu, ``Multi-objective optimization for {UAV} swarm-assisted {IoT} with virtual antenna arrays,'' \emph{{IEEE} Trans. Mob. Comput.}, vol.~23, no.~5, pp. 4890--4907, 2024.

\bibitem{Sun2021}
G.~Sun, J.~Li, Y.~Liu, S.~Liang, and H.~Kang, ``Time and energy minimization communications based on collaborative beamforming for {UAV} networks: A multi-objective optimization method,'' \emph{{IEEE} J. Sel. Areas Commun.}, vol.~39, no.~11, pp. 3555--3572, 2021.

\bibitem{alemdar2021rfclockAF}
K.~Alemdar, D.~Varshey, S.~Mohanti, U.~Muncuk, and K.~R. Chowdhury, ``Rfclock: timing, phase and frequency synchronization for distributed wireless networks,'' in \emph{Proc. {ACM} MobiCom}, 2021, pp. 15--27.

\bibitem{Mohanti2022}
S.~Mohanti, C.~Bocanegra, S.~G. Sanchez, K.~Alemdar, and K.~R. Chowdhury, ``{SABRE}: Swarm-based aerial beamforming radios: Experimentation and emulation,'' \emph{{IEEE} Trans. Wirel. Commun.}, vol.~21, no.~9, pp. 7460--7475, 2022.

\bibitem{Li2023TON}
J.~Li, G.~Sun, H.~Kang, A.~Wang, S.~Liang, Y.~Liu, and Y.~Zhang, ``Multi-objective optimization approaches for physical layer secure communications based on collaborative beamforming in {UAV} networks,'' \emph{{IEEE/ACM} Trans. Netw.}, vol.~31, no.~4, pp. 1902--1917, 2023.

\bibitem{IRS_sum_simple_method}
Z.~Wei, Y.~Cai, Z.~Sun, D.~W.~K. Ng, J.~Yuan, M.~Zhou, and L.~Sun, ``Sum-rate maximization for {IRS}-assisted {UAV} {OFDMA} communication systems,'' \emph{{IEEE} Trans. Wirel. Commun.}, vol.~20, no.~4, pp. 2530--2550, 2021.

\bibitem{li2020reconfigurableIRS-classic_los-channel}
S.~Li, B.~Duo, X.~Yuan, Y.~Liang, and M.~D. Renzo, ``Reconfigurable intelligent surface assisted {UAV} communication: Joint trajectory design and passive beamforming,'' \emph{{IEEE} Wirel. Commun. Lett.}, vol.~9, no.~5, pp. 716--720, 2020.

\bibitem{HanUAVIRSSecrecy}
S.~Han, J.~Wang, L.~Xiao, and C.~Li, ``Broadcast secrecy rate maximization in {UAV}-empowered {IRS} backscatter communications,'' \emph{{IEEE} Trans. Wirel. Commun.}, vol.~22, no.~10, pp. 6445--6458, 2023.

\bibitem{wang2025graph}
X.~Wang, L.~Feng, J.~Wang, H.~Du, C.~Zhao, W.~Li, Z.~Xiong, D.~Niyato, and P.~Zhang, ``Graph diffusion-based aebs deployment and resource allocation for rsma-enabled urllc low-altitude economy networks,'' \emph{arXiv preprint arXiv:2507.04081}, 2025.

\bibitem{Ma2024}
L.~Ma, Y.~Che, S.~Luo, and C.~M. Victor~Leung, ``Uav-aided 1d wireless power transfer with non-linear energy harvesting,'' in \emph{Proc. {IEEE/CIC} {ICCC}}, Aug. 2024, pp. 1513--1518.

\bibitem{haarnoja2018SAC}
T.~Haarnoja, A.~Zhou, P.~Abbeel, and S.~Levine, ``Soft actor-critic: Off-policy maximum entropy deep reinforcement learning with a stochastic actor,'' in \emph{Proc. {ICML}}, vol.~80, 2018, pp. 1856--1865.

\bibitem{liang2025rate}
S.~Liang, C.~Zhu, Z.~Yang, C.~You, D.~Niyato, K.-K. Wong, and Z.~Zhang, ``Rate maximization for fluid antenna system assisted semantic communication,'' \emph{arXiv preprint arXiv:2506.22943}, 2025.

\bibitem{vaswani2017attention}
A.~Vaswani, N.~Shazeer, N.~Parmar, J.~Uszkoreit, L.~Jones, A.~N. Gomez, L.~Kaiser, and I.~Polosukhin, ``Attention is all you need,'' in \emph{Proc. {NeurIPS}}, 2017, pp. 5998--6008.

\bibitem{iqbal2019actorMAAS}
S.~Iqbal and F.~Sha, ``Actor-attention-critic for multi-agent reinforcement learning,'' in \emph{Proc. {ICML}}, vol.~97, 2019, pp. 2961--2970.

\bibitem{Mao2025}
X.~Mao, G.~Wu, M.~Fan, Z.~Cao, and W.~Pedrycz, ``{DL-DRL:} {A} double-level deep reinforcement learning approach for large-scale task scheduling of multi-uav,'' \emph{{IEEE} Trans Autom. Sci. Eng.}, vol.~22, pp. 1028--1044, 2025.

\bibitem{Cheng2025}
M.~Cheng, S.~He, M.~Lin, W.~Zhu, and J.~Wang, ``{RL} and {DRL} based distributed user access schemes in multi-{UAV} networks,'' \emph{{IEEE} Trans. Veh. Technol.}, vol.~74, no.~3, pp. 5241--5246, 2025.

\bibitem{pmlr-v80-haarnoja18bMADDPG}
R.~Lowe, Y.~Wu, A.~Tamar, J.~Harb, P.~Abbeel, and I.~Mordatch, ``Multi-agent actor-critic for mixed cooperative-competitive environments,'' in \emph{Proc. {NeurIPS}}, 2017, pp. 6379--6390.

\bibitem{SAC2}
T.~Haarnoja, A.~Zhou, K.~Hartikainen, G.~Tucker, S.~Ha, J.~Tan, V.~Kumar, H.~Zhu, A.~Gupta, P.~Abbeel, and S.~Levine, ``Soft actor-critic algorithms and applications,'' \emph{CoRR}, vol. abs/1812.05905, 2018.

\bibitem{wu2019beamforming}
Q.~Wu and R.~Zhang, ``Beamforming optimization for wireless network aided by intelligent reflecting surface with discrete phase shifts,'' \emph{{IEEE} Trans. Commun.}, vol.~68, no.~3, pp. 1838--1851, 2019.

\end{thebibliography}
\end{document}